\documentclass[prd,aps,tightenlines,nofootinbib]{revtex4}

\usepackage{graphicx}

\newcommand{\be}{\begin{equation}}
\newcommand{\ee}{\end{equation}}
\newcommand{\bea}{\begin{eqnarray}}
\newcommand{\beas}{\begin{eqnarray*}}
\newcommand{\eea}{\end{eqnarray}}
\newcommand{\eeas}{\end{eqnarray*}}
\newcommand{\ba}{\begin{array}}
\newcommand{\ea}{\end{array}}

\newcommand{\nonum}{\newline}

\def\ls{\mathrel{\lower4pt\vbox{\lineskip=0pt\baselineskip=0pt
           \hbox{$<$}\hbox{$\sim$}}}}
\def\gs{\mathrel{\lower4pt\vbox{\lineskip=0pt\baselineskip=0pt
           \hbox{$>$}\hbox{$\sim$}}}} 

\newcommand{\la}{\langle}
\newcommand{\ra}{\rangle}

\def\nt{\not\hspace*{-.5ex}}

\begin{document}

\title{An Introduction to Extra Dimensions 
\footnote{Lectures given at the XI
Mexican School of Particles and Fields. Xalapa, M\'exico, August 1-13, 2004.}}

\author{Abdel P\'erez-Lorenzana}
\affiliation
{Departamento de F\'{\i}sica,
 Centro de Investigaci\'on y de Estudios 
Avanzados del I.P.N. \\
Apdo. Post. 14-740, 07000, M\'exico, D.F., M\'exico }


\begin{abstract}
Models  that involve extra dimensions have 
introduced completely new ways of looking up on old problems
in theoretical physics.  
The aim of the present notes is to provide a brief  introduction
to the  many uses that extra dimensions have found over the last few years, 
mainly following  an effective field theory point of view. 
Most parts of the discussion are
devoted to models with flat extra dimensions, covering both theoretical and 
phenomenological aspects. We also discuss some of the new 
ideas for model building where extra dimensions may play a role, 
including symmetry breaking by diverse new and old mechanisms.
Some interesting applications of these ideas are discussed over the 
notes, including models for neutrino masses and proton stability.
The last part of this review addresses some aspects 
of warped extra dimensions, and graviton localization.
\end{abstract}

\date{February, 2005}
\maketitle 


\section{Introduction: Why Considering Extra Dimensions?}
Possible existence of new spatial dimensions beyond the  four we see
have been under consideration for about eighty years already. 
The first ideas date back to the early 
works of Kaluza and Klein around  the 1920's~\cite{kk}, 
who tried to unify electromagnetism with Einstein gravity by 
proposing a theory with a compact fifth dimension, where
the photon was originated from the extra components of the metric.
In the course of the last few years there has been 
some considerable activity in 
the study of models that involve new extra spatial dimensions, 
mainly motivated  from theories that try to
incorporate gravity and  gauge interactions in a unique scheme  in
a reliable manner.
Extra dimensions are indeed a known fundamental ingredient 
for String Theory, since all versions of the theory are 
naturally and consistently formulated only in a space-time 
of more than four dimensions  (actually 10, or 11 if there is M-theory).
For some time,  however, it was conventional to assume that such
extra dimensions were compactified to manifolds of small radii, 
with sizes about the order of the  Planck length, 
$\ell_P\sim ~ 10^{-33}$~cm, 
such that they would remain hidden to the experiment, 
thus explaining why we see only four  dimensions.
In this same pictures, it was believed that the relevant energy 
scale where quantum gravity (and stringy)
effects would become important is given by the Planck mass,
which is  defined through the fundamental constants, 
including gravity Newton constant, as
\be 
  M_P c^2 = \left[\frac{\hbar c^5}{8\pi G_N}\right]^{1/2} ~\sim~  
   2.4\times 10^{18}~{\rm GeV}~;
   \label{dmp}
\ee
from where one defines $\ell_P = \hbar /M_P~c$. 
It is common to work in  natural units system which take  
$c=1=\hbar$, such that 
distance and time are measured in inverse units of energy. 
We will do so hereafter, unless otherwise stated. 
Since $M_P$ is quite large, 
there was little hope for experimentally probing such a regime, 
at least in the near future. 

From the theoretical side, this point of view
yet posses a fundamental puzzle, 
related to the quantum instability of the Higgs vacuum that fixes the
electroweak scale around $m_{EW}\sim 1~TeV$. 
Problem is that from  one loop order corrections, 
using a cut-off  regularization, one gets
bare mass independent 
quadratic divergences for the physical Higgs mass:
 \be
 \delta m_H^2
 ={1\over 8\pi^2}\left(\lambda_H-\lambda_F^2\right) \Lambda^2
 + (\mbox{log. div.})
 + \mbox{finite terms.}
 \label{quads}
 \ee
where $\lambda_{H}$ is the self-couplings of the Higgs field $H$, and 
$\lambda_{F}$ is the Yukawa coupling  to fermions. 
As the natural  cut-off $\Lambda$ of the theory is usually
believed to be the Planck scale or the GUT
scale, $M_{GUT}\sim 10^{16}~GeV$, 
this means that in order to get $m_H^2\sim m_{EW}^2$ we require to adjust 
the counterterm to at least one part in $10^{15}$.
Moreover, this adjustment must be made at each order in perturbation
theory. This large fine tuning is what  is known as the  hierarchy
problem.
Of course, the quadratic divergence can be renormalized away in exactly
the same manner  as it is done for logarithmic divergences, and in principle,
there is nothing formally wrong with this fine tuning. 
In fact,  if this calculation is performed  in the
dimensional regularization scheme, $DR$, one obtains only $1/\epsilon$
singularities which are absorbed into the definitions of the counterterms,
as usual.  Hence, the problem of quadratic divergences does not become
apparent there.  It arises  only when one attempts to give a physical
significance to the cut-off $\Lambda$. In other words, if the SM were a
fundamental theory then  using $DR$ would be justified. However, 
most theorists believe that the final theory should also include gravity,
then a cut-off must be introduced in the SM, regarding 
this fine tuning as unattractive. 
Explaining the hierarchy problem  has been a leading motivation to
explore new physics during the last twenty years, including
Supersymmetry~\cite{susy} and compositeness~\cite{tech}.

Recent developments,  based on the studies of
the non-perturbative regime of the $E_8\times E_8$ theory by Witten and 
Horava~\cite{witten}, have suggested that some, 
if not all, of the extra dimensions  could rather be larger than $\ell_P$.
Perhaps motivated by this, some authors  started to ask the question of how
large could these extra dimensions  be without getting into conflict with
observations, and even more interesting, where and how would this extra
dimensions manifest themselves.  
The intriguing answer to the first question 
point towards  the possibility that extra dimensions as large as
millimeters~\cite{dvali} could exist  and yet  remain hidden to the
experiments~\cite{expt,giudice,lykken,coll1,sn87,coll2}.  
This would be possible if our observable world is 
constrained to live on a four dimensional hypersurface (the brane) 
embedded in a higher dimensional space (the bulk), 
such that the extra dimensions can only be tested by gravity, 
a picture that resembles  D-brane theory constructions. 
Although it is fair to say that  similar ideas  were already 
proposed on the 80's by several authors~\cite{rubakov},
they were missed by some time, until the 
recent developments on string theory provided an independent realization
of such models~\cite{witten,anto,marchesano,lykk,dbranes}, given them certain credibility.
Besides, it was also the intriguing observation 
that such large extra dimensions would accept a scale of 
quantum gravity  much smaller than $M_P$, even closer to $m_{EW}$, 
thus offering an alternative solution to the hierarchy problem, 
which  attracted the attention of the community  towards this ideas.

To answer the second question many
phenomenological studies have been done, 
often based on simplified field theoretical models that are 
built up on a bottom-up approach, using an effective field theory 
point of view, with out almost any real string theory calculations.
In spite of being  quite speculative, and 
although it is unclear whether any of those models
is realized in nature, they still might provide some insights
to the low energy manifestations of the fundamental theory, 
since it is still possible that the excited modes of the string could appear
on the experiments way before any  quantum gravity effect, in which
case the effective field theory approach would be acceptable.

The goal for the  present notes is  to provide a general and brief introduction
to the field for the beginner. Many variants of the very first scenario
proposed  by Arkani-Hammed, Dimopoulos and D'vali (ADD)~\cite{dvali} have been
considered over the years, and there is no way we could comment all. Instead,
we shall rather concentrate on some of the most common aspects shared by those
models. This, in turn,  will provide us with the insight to extend our study to
other more elaborated ideas for the use of extra dimensions. 

The first part of these notes will  cover the basics of the ADD model.
We shall start discussing how the fundamental gravity scale departs from $M_P$
once extra dimensions are introduced, and the determination of the effective  
gravity coupling.
Then, we will introduce the basic field theory prescriptions used to construct
brane models and  
discuss the concept of dimensional reduction  on  compact
spaces and the resulting Kaluza-Klein (KK) mode expansion of bulk fields, which
provide the effective four dimensional  theory on which most calculations are
actually done.  We use these concepts to address some aspects of graviton
phenomenology  and briefly discuss some of the phenomenological bounds   for
the size of the extra dimensions and the fundamental gravity scale.

Third section is devoted to present some general aspects of the use of extra
dimensions in model building. Here we will review the KK decomposition for
matter and gauge fields, and discuss the concept of universal extra dimensions.
We will also address the possible phenomenology that may come 
with KK matter and gauge fields, with particular interest on 
the  power law running effect on gauge couplings. 
Our fourth section intends to be complementary to the third one. 
It provides a short review  on  many new ideas 
for the use of extra dimensions on the breaking of symmetries. 
Here we include spontaneous breaking on the bulk; shinning mechanism; 
orbifold breaking; and Scherk-Schwarz mechanisms.

As it is clear, with a low fundamental scale, as pretended by the ADD model, 
all the particle physics phenomena that usually invokes high energy scales does 
not work any more. Then, standard problems as  gauge coupling unification,  the
origin of neutrino masses and mixings and   proton decay; 
should be reviewed. Whereas the first  point is being already addressed 
on the model building section, 
we dedicate our fifth section to discuss some interesting ideas
to control lepton and baryon number violation in the context of 
extra dimension models. Our discussion includes a series of examples for
generating neutrino masses in models with low fundamental scales which make use
of bulk fields. We also address proton decay in the context of six dimensional
models where orbifold spatial symmetries provide the required control of this
process. The concept of fermion wave function localization 
on the brane is also discussed.

Finally, in section six we focus our  interest  on  
Randall and Sundrum models~\cite{merab,rs1,rs2,nimars} 
for warped backgrounds, for both compact and  infinite extra dimensions.
We will show in detail
how these solutions arise, as well as how gravity behaves in such theories. 
Some further ideas that include stabilization of the extra dimensions 
and graviton localization at branes are also covered.

Due to the nature of these notes, many other topics  are not
being covered, including brane intersecting models,  
cosmology of models with extra
dimensions both in flat and warped bulk backgrounds; KK dark matter; 
an extended discussion on black hole physics; as well as many
others.  The interested reader that would
like to go beyond the present notes can consult any of the 
excellent reviews that are now in the literature for references, 
some of which are given in references~\cite{reviews,review1}. 
Further references are also given at the conclusions.

\section{Flat and Compact Extra Dimensions: ADD Model}
 \subsection{Fundamental vs. Planck scales}

The existence of more than four dimensions in nature,  even if they
were small, may  not be completely harmless. They could have 
some visible manifestations in our (now effective) four dimensional world. 
To see this,  one has first to understand how the effective four
dimensional theory arises from the higher dimensional one. 
Formally,  this can be achieve by dimensionally reducing the complete theory, 
a concept that  we shall further  discuss  later on. 
For the moment, we must remember that gravity is a geometric property of the
space. Then, first thing to notice is that  in a higher dimensional world,
where Einstein gravity is assumed to hold, the  gravitational coupling does not
necessarily coincide with the well known Newton constant $G_N$, which is,
nevertheless, the gravity coupling we do observe. 
To explain this more clearly, let us assume as in Ref.~\cite{dvali}
that there are $n$   extra space-like dimension which are compactified
into circles of the  same radius R 
(so the space is factorized as a ${\cal M}_4\times T^n$ manifold).
We will call the fundamental gravity coupling $G_\ast$, and then write down the
higher dimensional gravity action:
 \be 
 S_{grav} = -\frac{1}{16\pi G_\ast} \int \! d^{4+n} x
 ~\sqrt{|g_{(4+n)}|}~ R_{(4+n)}~; 
 \label{sgrav}
 \ee
where $g_{(4+n)}$ stands for the metric in the whole $(4+n)$D space, 
$ds^2 = g_{MN}dx^Mdx^N$,
for which we will always use the  $(+,-,-,-,\dots)$ sign convention, and 
$M,N = 0,1,\dots, n +3$. 
The above action has to have proper dimensions, 
meaning that the extra length dimensions that come from the extra volume 
integration have to be equilibrated
by the dimensions on the gravity coupling. 
Notice that in natural units $S$ has no dimensions, whereas if 
we  assume for simplicity that the metric  $g_{(4+n)}$ is being 
taken dimensionless, so $[R_{(4+n)}] = [Length]^{-2} = [Energy]^2$.
Thus, the fundamental gravity coupling has to have dimensions
$[G_\ast] = [Energy]^{-(n+2)}$. In contrast, for the Newton constant, we have 
$[G_N] = [Energy]^{-2}$.
In order to extract the four dimensional gravity action let us  assume
that the extra dimensions are flat, thus, the metric has the form 
   \be 
     ds^2 = g_{\mu\nu}(x) dx^\mu dx^\nu - \delta_{ab}dy^a dy^b,
   \ee
where $g_{\mu\nu}$ gives the four dimensional part of the metric 
which depends only in the four dimensional coordinates $x^\mu$, 
for $\mu=0,1,2,3$; and 
$\delta_{ab}dy^a dy^b$ gives the line element on the bulk, 
whose coordinates are parameterized by $y^a$, $a=1,\dots,n$.
It is now easy
to see that $\sqrt{|g_{(4+n)}|}= \sqrt{|g_{(4)}|}$ and 
$R_{(4+n)} = R_{(4)}$, so one can integrate out the extra
dimensions in Eq.~(\ref{sgrav})  to get the effective 4D action
  \be 
  S_{grav} = -\frac{V_n}{16\pi G_\ast} \int \! d^{4}x 
  ~\sqrt{|g_{(4)}|}~ R_{(4)}~; 
  \label{eq4}
  \ee 
where $V_n$ stands for the volume of the extra space. For the torus
we simply take $V_n=R^n$.
Equation (\ref{eq4}) is precisely the 
standard gravity action in 4D if one makes the
identification, 
  \be 
    G_N = G_\ast/V_n~.
  \label{add0}
 \ee 
Newton constant is therefore given by a volumetric scaling of the truly
fundamental gravity scale.  Thus, $G_N$ is in fact an effective quantity.
Notice that even if $G_\ast$ were a large coupling (as an absolute number), 
one can still understand the smallness of $G_N$ 
via the volumetric suppression. 

To get a more physical meaning of these observations, let us consider a simple
experiment. Let us assume a couple of particles of masses $m_1$ and 
$m_2$, respectively, located on the hypersurface $y^a = 0$, and separated from 
each other  by a  distance $r$. The gravitational flux among both
particles would spread all over  the whole $(4+n)$~D space, however, since
the extra dimensions are compact,  the effective strength of the gravity
interaction would have two clear limits: 
(i) If both  test particles are
separated from each other  by a  distance $r \gg R$, 
the torus would effectively 
disappear for the four dimensional observer, the gravitational flux then gets
diluted by the extra volume and the observer would
see the  usual (weak) 4D gravitational potential
 \be 
 U_N(r) = -G_N\frac{m_1 m_2}{r} \label{ur1}~.
 \ee
(ii) However, if $r \ll R$, the 4D  observer would be able to
feel the presence of the bulk through  
the  flux that goes into the extra space, and 
thus, the  potential between each particle would appear to be stronger:
\be
U_\ast(r)= - G_\ast \frac{m_1 m_2}{ r^{n+1}}~.
\label{uast}
\ee
It is precisely the volumetric factor which does the matching 
of  both regimes of the theory. 
The change in the short distance behavior of the Newton's gravity law
should be observable in the experiments when measuring $U(r)$ for distances below
$R$. Current search for such deviations 
has gone down to 160 microns, so far with no signals of extra
dimensions~\cite{expt}.

We should now recall that the Planck scale, $M_P$,   
is defined in terms of the Newton constant, via Eq.~(\ref{dmp}).
In the present picture, it is then clear that $M_P$ is not fundamental anymore. 
The true scale for quantum gravity should rather 
be given in terms of $G_\ast$ instead.
So, we define the fundamental (string) scale as
  \be 
  M_\ast c^2=   \left[
  \frac{\hbar^{1+n} c^{5+n}}{8\pi G_\ast}\right]^{1/(2+n)}~,
  \ee
where, for comparing  to Eq. (\ref{dmp}), 
we have inserted back the corresponding  $c$ and $\hbar$ factors.
Clearly, coming back to  natural units, 
both scales are then related to each other by~\cite{dvali}
 \be
  M_{P}^2 = M_\ast^{n+2} V_n  ~.
  \label{mp}
 \ee
From  particle physics we already know that there is no evidence of 
quantum gravity (neither supersymmetry, nor string effects) well up to
energies around few hundred GeV, which says that $M_\ast\geq 1$~TeV. 
If the volume were large enough, 
then the fundamental scale could be as low as the electroweak scale, 
and there would be no hierarchy in the fundamental scales of physics, 
which so far has been considered as a puzzle. 
Of course, the price of solving the hierarchy problem this way 
would be now to explain why the extra dimensions are so large.
Using 
$V\sim R^n$ one can reverse above relation and get a feeling of the
possible values of $R$ for a given $M_\ast$. This is done 
just for our  desire of
having the quantum gravity scale as low as possible, perhaps to be accessible to
future experiments, although, the actual value is really unknown. 
As an example, if one takes $M_\ast$ to be 1~TeV; for
$n=1$, $R$ turns out to be about the size of the solar system 
($R\sim 10^{11}$~m)!, whereas for $n=2$ one gets $R\sim 0.2$~mm, that is
just at the current limit of short distance gravity experiments~\cite{expt}.
Of course, one single large extra dimension is not totally ruled out. 
Indeed, if one imposes the condition that $R<160 \mu m$ for $n=1$, 
we get  $M_\ast>10^8 GeV$.
More than two extra dimensions are in fact  expected
(string theory predicts six more), but in the final theory
those dimensions may turn out to have different sizes, or even geometries. 
More complex scenarios with a hierarchical distributions of the
sizes could be possible. 
For getting an insight of the theory, however, one usually  relies in toy
models with a single compact extra dimension,  implicitly 
assuming that the effects of the other compact dimensions 
do decouple from the effective theory.

It is worth mentioning that the 
actual  relationship among fundamental,  Planck, and compactification scales 
does depend on the type of metric one assumes for 
the compact space (although this  should be com-formally flat at least). 
This has inspired some variants on the
above simple picture. We will treat in 
some detail one specific example in the
last part of these notes,  the Randall-Sundrum models. 
There is also another nice example I would like to mention 
where $M_* R$ is rather of order one.
This is possible if extra dimensions are compactified on a 
hyperbolic manifold, where the volume is exponentially amplified from the 
curvature radius, which  goes
as  $R e^a$, with  the  constant $a$ depending
on the pattern of compactification~\cite{hipered}. 
In these models it is possible to get large volumes
with generic values of $R$ by properly choosing the parameter $a$. 
In what follows, however, we will address only ADD models for simplicity.

\subsection{Brane and Bulk Effective Field Theory prescriptions} 

While submillimeter dimensions remain untested for gravity, the particle
physics forces have certainly been accurately measured  up to weak scale
distances (about $10^{-18}$~cm).   Therefore, the Standard Model (SM) 
particles can not freely propagate in those large extra dimensions,  but must
be constrained to live on  a four dimensional submanifold. Then the  scenario 
we have in mind is one where we live in a four dimensional surface embedded in
a higher dimensional space. Such a surface shall be called a ``brane'' (a short
name for membrane). This picture is similar to the D-brane
models~\cite{dbranes},  as in the Horava-Witten theory~\cite{witten}. We may
also imagine our world as a  domain wall of size $M_\ast^{-1}$  where the
particle fields are trapped by some dynamical mechanism~\cite{dvali}. Such
hypersurface or brane would then be located at an specific point on the extra
space, usually, at the fixed points of the compact manifold. Clearly, such
picture breaks translational invariance, which may be reflected in two ways in
the physics of the model, affecting the flatness of the extra space (which
compensates for the  required flatness of the brane), and introducing a source
of violation  of the extra linear momentum. First point would drive us to the
Randall-Sundrum Models, that we shall discuss latter on. Second point would be
a constant issue along our discussions.

What we have called a brane in our previous paragraph is actually an effective
theory description. We have chosen to think up on them as  
topological defects (domain walls) of almost zero width, 
which could have fields localized to  its surface.
String theory D-branes  (Dirichlet branes)
are, however,  surfaces where open string can end on. 
Open strings  give rise to all kinds of fields localized 
to the brane, including gauge fields. In the supergravity 
approximation these D-branes will also appear as solitons of the 
supergravity equations of motion. 
In our approach we shall care little about  
where these branes come from, and rather simply assume
there is some consistent high-energy theory,  that 
would give rise to these objects, and which should appear at the fundamental
scale $M_\ast$. Thus, the natural UV cutoff of our models would always be given
by the quantum gravity scale.

D-branes are usually characterized by the number of spatial dimensions on the
surface. Hence, a $p$-brane is described by a flat space time with 
$p$ space-like  and one  time-like coordinates.  
The simplest model, we just mentioned above ,
would consist of SM particles living on a 3-brane.
Thus, we need  to describe theories that live either on the brane 
(as the Standard Model) or in the bulk (like gravity and perhaps SM singlets), 
as well as the interactions among these two theories.
For doing so we use the following field theory prescriptions: 
\begin{itemize}
\item[(i)] Bulk theories are 
 described by the higher dimensional action, defined in terms of a 
Lagrangian density of bulk fields,  $\phi(x,\vec{y})$, valued  
on all the space-time coordinates of the  bulk 
 \be 
    S_{\rm bulk}[\phi] = \int \! d^4x\, d^n y~
    \sqrt{|g_{(4+n)}|}{\cal L}(\phi(x,\vec{y}))~,
   \label{sbulk}
 \ee
where, as before, $x$ stands for the (3+1) coordinates of the brane and 
$y$ for the $n$ extra dimensions. 
\item[(ii)] Brane theories are described by the (3+1)D action of the brane
fields, $\varphi(x)$, which is
naturally promoted into a  higher dimensional expression by the use of a 
delta density:
 \be 
    S_{\rm brane}[\varphi] = \int \! d^4x\, d^n y~
    \sqrt{|g_{(4)}|}{\cal L}(\varphi(x))~
    \delta^n({\vec y}-{\vec y}_0)~,
   \label{sbrane}
 \ee
where we have taken the brane to be 
located at the position ${\vec y}={\vec y}_0$
along the extra dimensions, and  $g_{(4)}$ stands for the (3+1)D induced
metric on the brane. Usually we will work on flat space-times, unless
otherwise stated.
\item[(iii)] Finally, the action may contain terms that couple bulk to brane
fields. Last are localized on the space, thus, 
it is natural that a delta density would be involved 
in   such terms, say for instance
 \be
     \int \! d^4x\, d^n y~ 
    \sqrt{|g_{(4)}|}~ \phi(x,\vec{y})\, \bar\psi(x)\psi(x)~
    \delta^n({\vec y}-{\vec y}_0)
   \label{bb0}
 \ee
\end{itemize}

 \subsection{Dimensional Reduction}
The presence of delta functions in the previous action terms 
does not allow for a transparent interpretation, 
nor for an easy reading out of the  theory  dynamics.  
When they are present it is more useful to work in the effective four
dimensional theory that is obtained after integrating over the extra
dimensions. This procedure is generically called dimensional reduction.
It also helps to identify the low energy limit of the theory (where the extra
dimensions are not visible).

{\it A 5D toy model.-} 
To get an insight of what the effective 4D theory looks like, 
let us consider a simplified five dimensional toy model where the
fifth dimension has been compactified on a circle of radius $R$. 
The generalization of these results to more dimensions would be 
straightforward. 
Let $\phi$ be  a  bulk scalar field  for which the action
on flat space-time has the form 
 \be  
 S[\phi] = \frac{1}{2}\int\! d^4x\ dy\  
 \left(\partial^A\phi\partial_A\phi  - m^2 \phi^2\right) ;
 \label{sp}
 \ee
where now $A=1,\dots,5$, and $y$ denotes the fifth dimension. 
The compactness of the internal manifold is reflected in the 
periodicity of the field,  
$\phi(y) = \phi(y+2\pi R)$, which allows for a  
Fourier expansion as
 \be
 \phi(x,y) = \frac{1}{\sqrt{2\pi R}}\phi_0(x) 
  +   \sum_{n=1}^\infty \frac{1}{\sqrt{\pi R}} 
  \left[ \phi_n(x)\cos\left(\frac{ny}{ R}\right) 
   +  \hat\phi_n(x)\sin\left(\frac{ny}{R}\right)\right]. 
 \label{kkexp}
 \ee  
The very first term, $\phi_0$, 
with no dependence on the fifth dimension is 
usually referred as the zero mode. Other Fourier modes,
$\phi_n$ and $\hat\phi_n$; are called
the excited or Kaluza-Klein (KK) modes.
Notice the different normalization on all the excited modes,
 with respect to the zero mode.  
Some authors prefer to use a complex $e^{iny/R}$ Fourier expansion instead, 
but the equivalence of the procedure should be clear.

By introducing
last expansion into the action (\ref{sp}) and integrating over the extra
dimension one gets
 \be 
 S[\phi] = \sum_{n=0}^\infty \frac{1}{2}\int\! d^4x 
 \left(\partial^\mu\phi_n\partial_\mu \phi_n - m_n^2 \phi_n^2\right)
 +  \sum_{n=1}^\infty \frac{1}{2}\int\! d^4x 
 \left(\partial^\mu\hat\phi_n\partial_\mu\hat\phi_n - 
 m_n^2\hat\phi_n^2\right),
 \ee
where the KK mass is given as $m_n^2 = m^2 + \frac{n^2}{R^2}$. Therefore,
in the effective theory, the higher dimensional field appears as an
infinite tower of  fields with masses $m_n$, with degenerated massive levels, 
but the zero mode, as depicted in figure 1. 
Notice that all excited  modes are fields with the same spin, and
quantum numbers as $\phi$.  They differ only in the KK number $n$,
which is also associated with the fifth component of the momentum, 
which is discrete due to compatification.
This can be also understood in general from the higher dimensional 
invariant $p^A p_A  = m^2$, which can be rewritten as the effective four
dimensional squared momentum invariant 
$p^\mu p_\mu = m^2 + {{\vec p}_\bot}\,^2$, 
where  ${\vec p}_\bot$ stands for the extra momentum components.

\begin{figure}[ht]
\includegraphics[width=18pc]{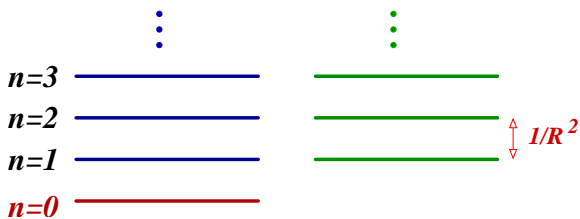}\hspace{2pc}%
\begin{minipage}[b]{16pc}
\caption{KK mass spectrum for a field on the circle.}
\end{minipage}
\end{figure}

Dimensionally reducing any higher dimensional field
theory (on the torus) 
would  give a similar spectrum for each particle with larger level degeneracy
($2^n$ states per KK level).
Different compactifications would lead to different mode expansions. 
Eq. (\ref{kkexp}) would had to be chosen  accordingly to the geometry of  the
extra space by typically using  
wave functions for free particles on such a 
space as the basis for the expansion.
Extra boundary conditions associated to specific topological properties of the
compact space may also help for a proper selection of the basis. 
A useful example is the one dimensional orbifold, $U(1)/Z_2$, 
which is built out of the circle, 
by identifying the opposite points around zero, so 
reducing the physical interval of the original circle to $[0,\pi]$ only. 
Operatively, this is done by requiring the theory to be invariant under the 
extra parity symmetry $Z_2: y\rightarrow -y$.
Under this symmetries all fields 
should pick up a specific parity, such that $\phi(-y)= \pm \phi(y)$. 
Even (odd) fields would then be expanded  only into cosine (sine)  modes, thus,
the KK spectrum would have only half of the modes (either the left or the
right tower in figure 1).
Clearly, odd fields do not have zero modes and thus do not appear at the low
energy theory.

For $m=0$, it is clear that
for energies below $\frac{1}{R}$ only the massless zero mode
will be kinematically accessible, making the theory looking four dimensional.
The appreciation of the impact of KK excitations
thus depends on the relevant energy of the experiment,
and on the compactification scale $\frac{1}{R}$:
\begin{itemize}
\item[(i)]
For energies $E\ll \frac{1}{R}$ 
physics would  behave purely as four dimensional.
\item[(ii)] At larger energies, $\frac{1}{R}<E<M_\ast$,
 or equivalently as we do measurements at shorter
distances, a large number of  KK excitations, 
$\sim (ER)^n$, becomes kinematically accessible, 
and their contributions relevant for the physics. 
Therefore, right above the threshold of the first excited level, 
the manifestation of the KK modes will start evidencing  the higher
dimensional nature of the theory. 
\item[(iii)]
At energies above $M_\ast$, however,  our effective approach has to be 
replaced by the use  of the fundamental theory that describes 
quantum gravity phenomena.
\end{itemize}

{\it Coupling suppressions.-}
Notice that the five dimensional scalar field $\phi$ 
we just considered  has mass 
dimension $\frac{3}{2}$, in natural units. 
This can be easily seeing from the kinetic part 
of the Lagrangian, which involves two
partial derivatives with mass dimension one each, 
and the fact that the action is dimensionless. 
In contrast, by similar arguments, all excited modes have mass 
dimension one, which is consistent with the KK expansion (\ref{kkexp}). 
In general for $n$ extra dimensions we  get the mass dimension for 
an arbitrary field to be $[\phi] = d_4 + \frac{n}{2}$, 
where $d_4$ is the natural mass dimension of $\phi$ in four dimensions. 

Because this change on the dimensionality of
$\phi$, most interaction terms on the Lagrangian 
(apart from the mass term) 
would  all have dimensionful couplings.
To keep them dimensionless a mass parameter should be  introduced to
correct the dimensions.
It is common to use as the natural choice for this parameter
the cut-off of the theory, $M_\ast$. 
For instance, let us consider the quartic couplings of $\phi$ in 5D. 
Since all potential terms should be of dimension
five, we should write down $\frac{\lambda}{M_\ast} \phi^4$, with $\lambda$
dimensionless.  
After integrating the fifth dimension, this operator will generate quartic 
couplings among all KK modes. 
Four normalization factors containing
$1/\sqrt{R}$ appear in the expansion of $\phi^4$. Two of them will be 
removed by the integration, thus,  we are left with the  effective
coupling $\lambda/MR$.  By introducing Eq. (\ref{mp}) we  observe that the
effective couplings have the form
 \be 
 \lambda\left( \frac{M_\ast}{M_{P}}\right)^2 \phi_k\phi_l\phi_m\phi_{k+l+m}~;
 \ee
where the indices are arranged to respect the conservation of the fifth
momentum.  From the last expression we conclude that in the low energy theory
($E<M_\ast$),  even at the zero mode level,  the effective coupling appears
suppressed respect to the bulk theory.  Therefore, the  effective four
dimensional theory would be  weaker interacting compared to the bulk
theory.   Let us recall that  same  happens to gravity on the bulk, 
where  the 
coupling constant is stronger than the effective 4D coupling, due to the 
volume suppression given in Eq.~(\ref{add0}), or equivalently in
Eq.~(\ref{mp}). 

Similar arguments apply in general for brane-bulk couplings.
Let us,  for instance, consider the case of a brane fermion, $\psi(x)$,
coupled to our bulk scalar $\phi$ field. 
For simplicity we assume that the brane is located at the position $y_0=0$, 
which in the case of orbifolds corresponds  to a fixed point. 
Thus, as the part  of the action that describes the brane-bulk coupling 
we choose  the term
 \be
 \int\! d^4x\ dy\ \frac{h}{\sqrt{M_\ast}}
 \bar\psi(x)\psi(x)\phi(x,y=0)\ \delta(y)
 = \int\! d^4x \frac{M_\ast}{M_{P}}h \cdot \bar\psi\psi
 \left(\phi_0 +  \sqrt{2}\sum_{n=1}^\infty  \phi_n\right).
 \label{Lint}
 \ee
Here the Yukawa coupling constant $h$ is dimensionless and the suppression
factor $1/\sqrt{M}$ has been introduce to correct the dimensions. 
On the right hand
side we have used the expansion (\ref{kkexp}) and Eq. (\ref{mp}). From
here, we notice that the coupling of brane to bulk fields is generically
suppressed by the ratio $M\over M_{P}$. Also, notice that   the modes
$\hat\phi_n$  decouple from the brane. Through this coupling we could not
distinguish the circle from the orbifold compactification.

Let us stress that the couplings in Eq. (\ref{Lint}) do not conserve  the KK
number. 
This reflects the fact that the   brane  breaks
the translational symmetry along the extra dimension. Nevertheless, it is
worth noticing that the four dimensional  theory is still Lorentz
invariant. Thus, if we reach enough energy on the brane, 
on a collision for instance,
as  to produce real emission of KK
modes, part of the energy of the brane would be  released into the bulk.
This would be the case of gravity, since the graviton is naturally a field that
lives in all dimensions.

Next, let us consider the scattering process among brane fermions 
$\psi\psi\rightarrow\psi\psi$ 
mediated by all the KK excitations of some field $\phi$. A  typical
amplitude will receive the contribution
 \be
 {\cal M} = \hat h^2  
\left( \frac{1}{ q^2 - m^2} +  2\sum_{n=1}\frac{1}{q^2 - m_n^2}\right) D(q^2) ,
 \ee 
where $\hat h = (M_\ast/M_P)h$ 
represents the effective coupling, and $D(q^2)$ is an
operator that only depends on the 4D Feynman rules of the involved fields.
The sum can easily be performed in this simple case, and one gets
 \be
 {\cal M} = \frac{ \hat h^2 \pi R}{\sqrt{q^2 - m^2}}
 \cot\left[\pi R   \sqrt{q^2 - m^2}\right] D(q^2).
 \ee
In more than five dimensions the equivalent to the above sum usually
diverges and has  to be regularized by introducing a cut-off at the
fundamental scale.  

We can also consider some simple limits to get a better feeling on the 
KK contribution to the process. At low energies, for instance, 
by assuming that $q^2\ll m^2\ll 1/R^2$
we may integrate out all the KK excitations, and at the
first order we get the amplitude
 \be
 {\cal M}\approx {\hat h^2\over m^2}\left(1+{\pi^2\over 3}m^2R^2\right) D(q^2).
 \label{msum}
 \ee
Last term between parenthesis is a typical effective 
correction produced by the
KK modes exchange to the pure four dimensional result.

On the other hand, at high energies, $qR\gg 1$, the
overall factor becomes $\hat h^2 N$, where $N = MR= M_{P}^2/M_\ast^2$ 
is the number of KK modes up to the cut-off. 
This large number of modes would overcome the suppression in the effective
coupling, such that one gets the amplitude ${\cal M}\approx h^2D(q^2)/q^2$, 
evidencing the 5D nature of the theory,  where  there is actually just 
a single higher dimensional 
field being exchange but with a larger coupling.

\subsection{Graviton Phenomenology and  Some Bounds}
\subsubsection{Graviton couplings and the effective gravity
interaction law.}

One of the first physical examples of a brane-bulk interaction one may be 
interested in analyzing with some  care is the effective 
gravitational coupling of particles  located at the brane, 
for which one  needs to understand the way gravitons couple to brane fields.
The problem has been extendedly discussed  by Giudice, Ratazzi and 
Wells~\cite{giudice}
and independently by Han, Lykken and Zhang~\cite{lykken} assuming a flat bulk. 
Here we summarize some of the main points.
We start from the action that describes a particle on the brane
 \be 
 S = \int \!d^4x d^ny\,\sqrt{|g(y^a=0)|}~{\cal L}~ \delta^{(n)}(y) ~,
 \ee
where the induced metric $g(y^a=0)$ 
now includes  small metric fluctuations 
$h_{MN}$ over flat space, which are also called the graviton, such that 
 \be 
 g_{MN} = \eta_{MN} + \frac{1}{2M_\ast^{n/2+1}} h_{MN}. 
 \ee
The source of those fluctuations are of course the energy on the brane, i.e.,
the matter 
energy momentum tensor $\sqrt{g}~ T^{\mu\nu} = \delta S/\delta g_{\mu\nu}$ 
that enters on the RHS of Einstein equations:
\[R_{MN} - \frac{1}{2} R_{(4+n)}g_{MN} = -\frac{1}{M_\ast^{2+n}}
T_{\mu\nu}\eta^{\mu}_M \eta^\nu_N \delta^{(n)}(y)~.\] 
The effective coupling, at first order in $h$, of matter to graviton field is
then described by the action
 \be 
 S_{int} = \int\!d^4x \frac{h_{\mu\nu}}{M_\ast^{n/2 +1}}T^{\mu\nu} 
 \label{gravt}
 \ee
 
It is clear  from the effective four dimensional
point of view, that the fluctuations $h_{MN}$ would have different 4D Lorentz
components. (i) $h_{\mu\nu}$ clearly contains a 4D Lorentz tensor, the true 
four dimensional graviton. (ii) $h_{a\mu}$ behaves as a vector, the
graviphotons. (iii) Finally, $h_{ab}$  behaves as a group of scalars
(graviscalar fields), one of which corresponds to the partial trace of $h$ 
(${h^a}_a$) that  we will call the radion field.
To count the number of degrees of freedom in $h_{MN}$ 
we should first note that $h$ is a $D\times D$ symmetric tensor, for
$D=4+n$. 
Next,  general coordinate invariance of general relativity can be translated
into $2n$ independent gauge fixing conditions, half usually chosen as the
harmonic gauge  $\partial_M h_N^M = \frac{1}{2} \partial_N h^M_M$. 
In total there are $n(n-3)/2$ independent degrees of freedom. Clearly,
for $n=4$ one has the usual two helicity states of a 
massless spin two particle. 

All those effective fields would of course  have a KK decomposition,
 \be 
 h_{MN}(x,y) = \sum_{\vec{n}} \frac{h_{MN}^{(\vec{n})}(x)}{\sqrt{V_n}}~
 e^{i\vec{n}\cdot\vec{y}/R}~,
 \ee
where we have assumed the compact space to be a torus of unique radius $R$,
also here $\vec{n} = (n_1,\dots,n_n)$, with all $n_a$ integer numbers.
Once we insert back the above expansion into $S_{int}$, it is not hard to see
that the volume suppression will exchange  the  $M_\ast^{n/2 +1}$ by an $M_P$ 
suppression for the the effective interaction with a single KK mode.
Therefore, all modes  couple with  standard gravity strength.
Briefly, only the 4D gravitons, $G_{\mu\nu}$, 
and the radion field, $b(x)$, get couple at
first order level to the brane energy momentum tensor~\cite{giudice,lykken}
 \be 
 {\cal L} = -\frac{1}{M_P}\sum_{\vec{n}} \left[G^{(\vec{n})\mu\nu} - 
 \frac{1}{3}\sqrt{\frac{2}{3(n+2)}}b^{(\vec{n})}\eta^{\mu\nu}\right] T_{\mu\nu}.
 \label{leff0}
 \ee
Notice that $G^{(0)\mu\nu}$ is
massless since the higher dimensional graviton $h_{MN}$ has no mass itself. 
That is the source of long range four dimensional gravity interactions. 
It is worth saying that on the contrary $b^{(0)}$ 
should not be  massless, otherwise it should  violate 
the equivalence principle,
since it would mean a scalar (gravitational) interaction of  long range too. 
$b^{(0)}$ should get a mass
from the stabilization mechanism that keeps the extra volume finite. 

Now that we know how gravitons couple
to  brane matter 
we can use this effective field theory point of view to calculate 
what the effective gravitational interaction law should be on the brane. 
KK gravitons are  massive, thus, 
the interaction mediated by them on the brane is of short range. 
More precisely, each KK mode contribute to the gravitational potential among
two test particles of masses $m_1$ and $m_2$ located on the brane, separated
by a distance $r$, with a Yukawa potential  
 \be 
 \Delta_{\vec{n}}U(r) \simeq -G_N \frac{m_1 m_2}{r} e^{-m_{\vec{n}} r}
  = U_N(r) e^{-m_{\vec{n}} r}~.
 \label{yukp}
 \ee
Total contribution of all KK modes, the sum over all KK masses
$m_{\vec{n}}^2 = \vec{n}^2/R$, can be estimated 
in the  continuum  limit, to get 
 \be 
 U_T(r) \simeq - G_N V_n (n -1)!~ \frac{m_1 m_2}{r^{n+1}}
 \simeq U_\ast(r)~,
 \ee
as mentioned in Eq.~(\ref{uast}).
Experimentally, however, for $r$ just around the threshold 
$R$ only the very first excited
modes would be relevant, and so, 
the potential one should see in short distance
tests of Newton's law should rather be of the form~\cite{sfetsos} 
 \be 
 U(r)\simeq U_N(r) \left( 1+ \alpha e^{-r/R} \right)~.
 \ee
where $\alpha=8n/3$  accounts for the multiplicity of the very first excited 
level. As already mentioned, recent 
measurements have tested inverse squared law of
gravity down to about 160 $\mu m$, and  no signals of deviation have 
been found~\cite{expt}.

\subsubsection{Collider physics.}

As gravity
may become comparable in strength to the gauge interactions at energies 
$M\ast \sim$ TeV, the nature of the  quantum theory of gravity would become
accessible to  LHC and NLC.
Indeed, the effect of the gravitational couplings would be mostly of
two types: (i) missing energy, that goes into the bulk; 
 and (ii) corrections to the
standard cross sections from graviton exchange~\cite{coll1}.
A long number of studies on this topics have 
appeared~\cite{giudice,lykken,coll1},  and 
some nice and short reviews of collider signatures
were early given in~\cite{review1}. 
Here we just briefly  summarize some of the possible signals. 
Some indicative bounds one can obtain from the experiments on the fundamental
scale are also shown in tables 1 and 2. 
Notice however that precise numbers do 
depend on the number of extra dimensions. 
At $e^+e^-$ colliders (LEP,LEPII, L3), the best signals
would be the production of gravitons with $Z,\gamma$ or fermion pairs
$\bar ff$.  In hadron colliders (CDF, LHC)
one could see graviton production in Drell-Yang
processes, and there is also the interesting  monojet
production~\cite{giudice,lykken} which is yet untested.  
LHC could actually impose bounds up to 4 TeV for $M_\ast$ for 10 $fb^{-1}$
luminosity. 

Graviton exchange  either leads to modifications of
the SM cross sections and asymmetries, or to new processes not allowed in
the SM at tree level. The amplitude for exchange of the entire tower
naively diverges when $n>1$ and has to be regularized, as already mentioned.  
An interesting channel is
$\gamma\gamma$ scattering, which appears at tree level, and may surpasses
the SM background at $s=0.5$ TeV for $M_\ast= 4$ TeV. 
Bi-boson productions of $\gamma\gamma$, $WW$ and $ZZ$ may also give some
competitive bounds~\cite{giudice,lykken,coll1}.
Some experimental limits, most of them based on existing data, are given
in Table 2. The upcoming experiments will 
easily overpass those limits. 

 \begin{table}[ht]
  \caption{Collider limits for the fundamental scale $M_\ast$. Graviton
 Production.}
 \begin{center}
 \begin{tabular}{llll}
\hline\hline
 Process & Background & $M_\ast$ limit & Collider \\
 \hline
 $e^+e^-\rightarrow \gamma G$ &  $e^+e^-\rightarrow\gamma\bar\nu\nu$
      & 1 TeV & ~L3 \\
 $e^+e^-\rightarrow ZG $ & $ e^+e^-\rightarrow Z\bar\nu\nu$ &
      $\bigg\{\begin{array}{l} 515~GeV \\ 600~GeV \end{array}$ &
      $\begin{array}{l} \mbox{LEPII} \\ \mbox{L3} \end{array}$ \\
 $Z\rightarrow \bar ffG$ &
 $Z\rightarrow  \bar ff\bar\nu\nu$ & 0.4 TeV &~LEP \\
\hline\hline
 \end{tabular}
 \end{center}
 \end{table}
 \begin{table}[ht]
 \caption{Collider limits for the fundamental scale $M_\ast$. Virtual Graviton
   exchange}
 \begin{center}
 \begin{tabular}{lll}
 \hline\hline
 Process  & $M_\ast$ limit & Collider \\
 \hline
 $e^+e^-\rightarrow f\bar f$ & 0.94 TeV &  ~Tevatron \& HERA\\
 $e^+e^-\rightarrow \gamma \gamma,WW, ZZ$ & 
       $\bigg\{\begin{array}{l}0.7 -- 1~TeV \\ 0.8~TeV\end{array}$&
	     $\begin{array}{l} \mbox{LEP}\\ \mbox{L3} \end{array}$ \\
 All above & 1 TeV & ~L3\\
  Bhabha scattering & 1.4 TeV & ~LEP \\
 $\begin{array}{l} q\bar q\rightarrow \gamma \gamma  \\
  gg\rightarrow \gamma \gamma  \end{array}\bigg\}$  & 0.9 TeV & ~CDF \\
\hline\hline
 \end{tabular}
 \end{center}
 \end{table}

Another intriguing phenomena in colliders, 
associated to a low gravity scale, is the possible production of 
microscopic Black Holes~\cite{bh1}. Given that the $(4+n)D$ Schwarzschild
radius 
 \be
 r_S \sim \left(\frac{M_{BH}}{M_\ast} \right)^{\frac{1}{1+n}}\frac{1}{M_\ast}
 \ee
may be larger than the impact parameter in a collision at energies larger than
$M_\ast$, it has been conjecture that a Black Hole may form with a mass 
$M_{BH} =\sqrt{s}$. Since the geometrical cross section of the Black Hole goes
as $\sigma\sim \pi r_S^2 ~~ \sim TeV^{-2}\approx 400~pb$, 
it has been pointed out
that LHC running at maximal energy 
could even be producing about $10^7$ of those Black Holes per year, 
if
$M_\ast\sim TeV$. However, such tiny objects are quite unstable. Indeed they 
thermally evaporate in a life time
 \be
 \tau\approx \frac{1}{M_\ast}\left(\frac{M_{BH}}{M_\ast}\right)^{(3n+1)/(n+1)}
 \ee
by releasing all its energy into Hawking radiation
containing predominantly brane modes. For above parameters one gets 
$\tau< 10^{-25}~sec$.
This efficient conversion of
collider energy into thermal radiation would be a clear signature of having
reached the quantum gravity regime. 

\subsubsection{Cosmology and Astrophysics.}

Graviton production 
may also posses strong constraints on the theory when considering that the
early Universe was an important resource of energy. 
How much of this energy could had 
gone into the bulk without affecting cosmological evolution? 
For large extra
dimensions, the splitting among two excited modes is pretty small, $1/R$. For
$n=2$ and $M_\ast$ at TeV scale this means a mass gap of just about
$10^{-3}$~eV. For a process  where the center mass energy is E, up to 
 $N= (ER)^n$ KK modes would be  kinematically accessible. During Big Bang
Nucleosynthesis (BBN), for instance, where $E$ was about few MeV,  this already
means more than $10^{18}$ modes for $n=2$.  
So many modes may be troublesome for a hot Universe that may 
release too much energy into gravitons. One can
immediately notice that the graviton creation rate, per unit time and volume, 
from  brane thermal processes at temperature $T$ goes as 
$$ \sigma_{total} = \frac{(TR)^n}{M_P^2} =
\frac{T^n}{M_\ast^{n+2}}. $$
The standard Universe evolution would be conserved as far as the total number
density of  produced 
KK gravitons, $n_g$, remains small when compared to 
photon number density, $n_\gamma$. This is a
sufficient condition that  can be translated into a bound for the
reheating energy, since as hotter the media as more gravitons can be excited.
It is not hard to see that this condition implies~\cite{dvali} 
 \be 
 \frac{n_g}{n_\gamma} \approx \frac{T^{n+1}M_P}{M_\ast^{n +2}} <1~.
 \ee
Equivalently, the maximal temperature our Universe could reach with out
producing to many gravitons must satisfy
 \be 
 T_r^{n +1}< \frac{M_\ast^{n+2}}{M_P}~.
 \label{trgt} 
 \ee
To give numbers consider for instance $M_\ast = 10$~TeV and $n=2$, 
which means  $T_r< 100~MeV$, just about to what is needed to have 
BBN working~\cite{steen} (see also \cite{hannestad}). 
The brane
Universe with large extra dimensions is then rather cold. This would be
reflected in some difficulties for those models trying to implement 
baryogenesis or leptogenesis based in electroweak energy physics.

 Thermal graviton 
emission is not restricted to early Universe. One can expect this to
happen in many other environments. We have already mention colliders as an
example. But even the hot astrophysical objects can be  graviton sources. 
Gravitons emitted by stellar objects take away energy, this
contributes to cool down the star. Data obtained  
from the supernova 1987a gives $M_\ast\gs 10^{\frac{15-4.5 n}{n+2}}$, 
which for $n=2$ means $M_\ast>30$ TeV~\cite{sn87}. 

Moreover, the Universe have been emitting gravitons all along its life. 
Those massive gravitons are actually unstable. They decay back into the brane
re-injecting  energy in the form of relativistic particles,
through channels like $G_{KK}\rightarrow \gamma\gamma;~ e^+e^-;~ \dots$, 
within a life time 
 \be 
 \tau_g\approx 10^{11}~{\rm yrs}~\times \left(\frac{30 MeV}{m_g}\right)^3~.
 \ee
Thus, gravitons with a mass about 30~MeV would be decaying at the present
time, contributing to the diffuse gamma ray background. EGRET and COMPTEL
observations
on this window of cosmic radiation do not see an important contribution, thus,
there could not be so many of such  gravitons decaying out there. 
Quantitatively it means that 
$M_\ast> 500$~TeV~\cite{hannestad,hannestad1}.

Stringent limits  come from the observation of neutron stars. 
Massive KK gravitons have small kinetic energy, so that a large fraction of
those produced in the inner supernovae core  remain gravitationally trapped. 
Thus, neutron stars would have a halo of KK gravitons, which is dark except
for the few MeV  $e^+e^-$ pairs and $\gamma$ rays produced by their
decay. Neutron stars are observed very close (as close as 60 pc), and so one
could observe this flux coming from the stars. GLAST, for instance, could be in
position of finding the KK signature, well up to a fundamental scale as
large as 1300 TeV for $n=2$~\cite{hannestad2}. 
Constraints from gamma ray emission from the whole population of 
neutron stars in the galactic bulge  against EGRET observations
gives limits on about $M_\ast > 450~TeV$ for $n=2$~\cite{casse}.

KK decay may also  heat the neutron star up to levels above the 
temperature expected from standard cooling models. 
Direct observation of neutron
star luminosity provides the most restrictive lower 
bound on $M_\ast$ at about 1700 TeV for $n=2$~\cite{hannestad2}. 
Larger number of dimensions
results in softer lower bounds since the  mass gap among KK modes increases.
These  bounds, however, depend on the graviton decaying
mainly into the visible Standard Model particles. Nevertheless, if
heavy KK gravitons decay into more stable lighter KK modes, 
with large kinetic energies, such bounds can be avoided, since 
these last KK modes would fly away from the star leaving no detectable 
signal behind.  
This indeed may happen if, for instance, translational invariance is broken in
the bulk, such that inter KK mode decay is not forbidden~\cite{nusinov}. 
Supernova cooling  and BBN bounds are, on the hand, more robust. 
 
Microscopic Black Holes may also be produced by 
ultra high energy  cosmic rays hitting the atmosphere, since these 
events may reach center mass energies well above $10^{5}$ GeV.  
Black Hole production, due to graviton mediated
interactions of ultra high energy neutrinos
in the atmosphere, would be manifested by deeply penetrating 
horizontal air showers~\cite{hecr}.
Provided, of course,  the fundamental scale
turns out to be at the TeV range. Auger, for instance, could  be
able to observe more than one of such events per year.
Contrary to earlier expectations,  some recent numerical simulations
have shown that black-hole induced air-showers do
not seem, however, to  possess a characteristic profile, 
and the rate of horizontal showers may not be higher than
in standard interactions~\cite{ave}. 

\section{Model Building}

So far we have been discussing the simple ADD model where 
all matter fields are assumed to live on the brane. 
However, there has been also  quite a large interest on the community  
in studying more complicated constructions where
other fields, besides gravity, live on more than four dimensions.
The first simple extension one can think 
is to assume than some other singlet fields may also
propagate in the bulk. 
These fields can either be scalars or fermions, and can be
useful for a diversity of new mechanisms. 
One more  step on this line of thought is to also  promote
SM fields to propagate in the extra dimensions. 
Although, this is indeed possible, some modifications have to be introduced
on the profile of compact space in order 
to control the spectrum and masses of KK excitations of SM fields.
These constructions contain a series of interesting properties
that may be of some use for model building, 
and it is  worth paying some attention to them. 

\subsection{Bulk Fermions} 
We have already discussed dimensional reduction with bulk scalar fields. 
Let us now turn our attention towards fermions. We start considering a massless
fermion, $\Psi$, in (4+n)D. Naively we will take it as 
the solution to the Dirac equation $i\partial_M\Gamma^M \Psi(x,y) = 0$ 
where $\Gamma_M$ satisfies the Clifford algebra
$\left\{ \Gamma^M,\Gamma^N\right\}= 2 \eta^{MN}$. The algebra 
now involves more gamma matrices than in four dimensions, 
and this have important implications on  degrees of
freedom of the spinors. Consider for instance the 5D case, 
where we use the representation
\be
 \Gamma^\mu = \gamma^\mu = 
 \left(\ba{cc} 0&\sigma^\mu\\ \bar\sigma^\mu & 0\ea\right); 
 \qquad \mbox{ and } \qquad
 \Gamma^4 = i \gamma_5 = \gamma^0\gamma^1\gamma^2\gamma^3 = 
   \left(\ba{cc} {\bf 1}&0\\ 0 & {\bf -1}\ea\right),
 \label{5dgammas}
 \ee
where as usual $\sigma^\mu = ({\bf 1},\vec{\sigma})$ and  
$\bar\sigma^\mu = ({\bf 1},-\vec{\sigma})$, with $\sigma_i$ 
the three Pauli matrices.
With $\gamma^5$ included among  Dirac matrices  
and because there is no any other matrix with the same properties of $\gamma_5$,
that is to say which  anticommutes with all $\gamma^M$ and  satisfies 
$(\gamma)^2 = {\bf 1}$,
there is not explicit chirality in the theory. 
In this basis, $\Psi$ is conveniently  written as
 \be 
 \Psi= \left(\ba{c} \psi_R\\ \psi_L\ea\right)~,
 \label{5dfer}
 \ee 
and thus  a 5D bulk fermion is necessarily a  four component spinor. 
This may be troublesome given that known four dimensional fermions are
chiral (weak interactions are different for left and right components),
but there are ways to fix this, as we will comment below.

Increasing even more the number of dimensions does not change this feature. For
6D there are not enough four by four anticommuting 
gamma matrices to satisfy the algebra, and one needs to go to larger matrices 
which can always be built out  of the same gamma matrices used for 5D.
The simplest representation is made of eight by eight matrices that 
be can be chosen as
\be
\Gamma^\mu = \gamma^\mu\otimes \sigma_1 =
            \left(\ba{cc}   
            0 & \gamma^\mu \\
            \gamma^\mu  & 0\\
              \ea\right)~;
\quad
\Gamma^4 = i\gamma_5\otimes \sigma_1 =
           \left(\ba{cc}   
            0 & i\gamma_5 \\
            i \gamma_5  & 0\\
               \ea\right)~;
\quad
\Gamma^5 = {\bf 1}\otimes i\sigma_2 =
           \left(\ba{cc}   
            0 & {\bf 1} \\
            -{\bf 1}  & 0\\
               \ea\right)~.
\label{6dgammas}
\ee
6D spinor would in general have eight components, but there is now  a 
$\Gamma^7 = \Gamma^0 \Gamma^1\cdots \Gamma^5 = 
{\rm diag}({\bf 1}, -{\bf 1})$ 
which anticommutes with all other gammas and satisfy $(\Gamma^7)^2 = 1$, 
thus one can define a 6D chirality in terms of the eigenstates of $\Gamma^7$,
however, the corresponding chiral states $\Psi_{\pm}$ are not equivalent to the
4D ones, they still are four component spinors, with both left and right
components as given in Eq.~(\ref{5dfer}).

In general, for $4+2k$ dimensions 
gamma matrices can be constructed in terms of
those used for $(4+2k-1)D$, following a similar prescription as the one used
above. In the simplest representation, for both $4+2k$ and
$4+2k+1$ dimensions they are squared matrices of
dimension $2^{k+2}$. In even dimensions ($4+2k$) we always have 
a $\bar\Gamma \propto \Gamma^0\cdots \Gamma^{3+2k}$
that anticommutes with all Dirac matrices in the algebra,
and it is such that  $(\bar\Gamma)^2=1$. 
Thus one can always introduce the concept of chirality associated to the
eigenstates of $\bar\Gamma$, but it does not correspond to the
known 4D chirality~\cite{sohnius}. In odd dimensions ($4+2k+1$)
one  may always choose
$\Gamma^{4+2k}= i\bar\Gamma$, and so, there is no chirality.

For simplicity let us now concentrate in the 5D case.
The procedure for higher dimensions should then  be straightforward.
To dimensionally reduce the theory we  start with the action for a massless
fermion, 
\be
S = \int\!d^4x dy\, i\bar\Psi\Gamma^A\partial_A\Psi = 
 \int\!d^4x dy\,  
 \left[   i\bar\Psi\gamma^\mu\partial_\mu\Psi + 
\bar\Psi\gamma^5\partial_y\Psi
 \right],
\label{sfer}
\ee
where we have explicitely used that  $\Gamma^4 = i\gamma^5$. 
Clearly, if one uses Eq.~(\ref{5dfer}), the last term on the RHS simply reads
$ \bar\psi_L\partial_y\psi_R -(L\leftrightarrow R)$.
Now we use the Fourier expansion for  compactification on the circle
\[\psi(x,y)  = 
\frac{1}{\sqrt{2\pi R}}\psi_0(x)
  +   \sum_{n=1}^\infty \frac{1}{\sqrt{\pi R}} 
  \left[ \psi_n(x)\cos\left(\frac{ny}{ R}\right)
  +  \hat\psi_n(x)\sin\left(\frac{ny}{R}\right)\right]~,
\]
where  $L,R$ indices on the spinors should be understood.
By setting  this expansion into the action it is easy to see that 
after integrating out the extra dimension,  the first term on the RHS of
Eq.~(\ref{sfer}) precisely gives the kinetic terms of all 
KK components, whereas  the last terms  become the KK Dirac-like mass terms:
 \be 
 \sum_{n=1}^\infty \int\!d^4x \left(\frac{n}{R}\right)
 \left[\bar\psi_{n,L}\hat\psi_{n,R} - (L\leftrightarrow R) \right]~.
 \ee
Notice that each of these terms couples even ($\psi_{n}$) to odd modes 
($\hat\psi_{n}$), and the two zero modes remain massless.
Regarding mass terms, two different Lorentz invariant fermion bilinears are possible in five
dimensions: Dirac mass terms $\bar\Psi\Psi$ and Majorana masses
$\Psi^TC_5\Psi$, where $C_5= \gamma^0\gamma^2\gamma^5$.
These terms do not give rise to mixing among even and odd KK modes, rather
for a 5D  Dirac mass term for instance, one gets 
 \be 
 \int\!dy\, m\bar\Psi\Psi = 
 \sum_{n=0}^\infty m\bar\Psi_n\Psi_n + 
 \sum_{n=1}^\infty m\hat{\bar\Psi}_n\hat{\Psi}_n~.
 \ee
5D Dirac mass, however,  is an odd function under the orbifold symmetry 
$y\rightarrow -y$, under which $\Psi\rightarrow \pm \gamma^5\Psi$, where the
overall sign remains as a free choice for each field.
So, if we use the orbifold $U(1)/Z_2$ instead of the circle for compactifying
the fifth dimension, 
this term should be zero. The orbifolding also takes care of the duplication of 
degrees of freedom. Due to the way $\psi_{L,R}$ transform, one of this
components becomes and odd field under $Z_2$ and therefore at zero mode level
the theory appears as if fermions were chiral.

\subsection{Bulk Vectors}

Lets now consider  a gauge field sited on five dimensions. 
For simplicity we will consider only the case of a free gauge abelian 
theory. The Lagrangian, as usual, is given as
 \be 
  {\cal L}_{5D} = -{1\over 4} F_{MN}F^{MN}  = 
  -{1\over 4} F_{\mu\nu}F^{\mu\nu} +  {1\over 2} F_{\mu5}F^{\mu5}\ ,
 \ee
where $F_{MN} = \partial_M A_N - \partial_N A_M$; 
and $A_M$ is the vector field
which now has an extra degree of freedom, $A_5$, 
that behaves as an scalar field in 4D.  
Now we  proceed as usual with the compactification of the theory, 
starting with the mode expansion
 \be 
 A_M(x,y)  =  \frac{1}{\sqrt{2\pi R}}A_M^{(0)}(x)
 + \sum_{n=1}\frac{1}{\sqrt{\pi R}}  
  \left[ A_M^{(n)}(x) \cos\left(\frac{ny}{R}\right) + 
  \hat A_M^{(n)}(x) \sin\left(\frac{ny}{R}\right)\right]~.
 \ee
Upon integration over the extra dimension one gets the 
effective Lagrangian~\cite{ddg}
 \be
 {\cal L}_{eff} = 
\sum_{n=0}\left\{  -\frac{1}{4}F_{\mu\nu}^{(n)}F^{\mu\nu}_{(n)} 
+ \frac{1}{2} \left[\partial_\mu A_5^{(n)} -
 \left(\frac{n}{R}\right)\hat A_\mu^{(n)}\right]^2
\right\}  +(A \leftrightarrow \hat A)~.
 \ee
Notice that the terms within squared brackets mix even and odd modes.  
Moreover, the Lagrangian contains a quadratic term in $A_\mu$, which then  
looks as a mass term. 
Indeed,  since  gauge invariance of the theory,
$A_M\rightarrow A_M + \partial_M \Lambda(x,y)$, 
can  also be expressed in terms of the (expanded) gauge 
transformation of the KK modes
\be
 A_\mu^{(n)} 
\rightarrow A_\mu^{(n)} + \partial_\mu \Lambda^{(n)}(x)~;
\qquad
A_5^{(n)}\rightarrow A_5^{(n)} + 
         \left(\frac{n}{R}\right) \hat\Lambda^{(n)}(x)~;
\ee
with similar expressions for odd modes.
We can use this freedom to diagonalize the mixed terms in the affective
Lagrangian by fixing the gauge.
 We simply take 
$\hat\Lambda^{(n)} = -(R/n)A_5^{(n)}$ 
[and $\Lambda^{(n)} = -(R/n)\hat A_5^{(n)}$] to get 
 \be
 {\cal L}_{eff}  = 
 \sum_{n=0}\left\{  -\frac{1}{4}F_{\mu\nu}^{(n)}F^{\mu\nu}_{(n)} 
 +\frac{1}{2}\left(\frac{n}{R}\right)^2 A_\mu^{(n)}A^\mu_{(n)} + 
 \frac{1}{2}\left(\partial_\mu A_5^{(0)}\right)^2\right\} 
 + (\mbox{odd modes})
 \ee
Hence, the only massless vector is the zero mode, all KK modes  acquire
a mass by absorbing the scalars $A_5^{(n)}$. This resembles the 
Higgs mechanism with $A_5^{(n)}$ playing the role of the 
Goldston bosons  associated to the  spontaneous
isometry breaking~\cite{dobado}.  Nevertheless,
there remain a massless U(1) gauge field and 
a massless scalar at the zero mode level, thus   
the gauge symmetry at this level
of the effective theory remains untouched. 
Once more, if one uses an orbifold, 
the extra degree of freedom, $A_5^{(0)}$, that appears at zero level can be
projected out. This is because under $Z_2$, $A_5$
can be chosen to be an odd function.

{\it Gauge couplings}.- We have already mention that, for bulk theories, the
coupling constants usually get a volume suppression that makes the theory 
looking weaker in 4D. With the gauge fields on the bulk one has  the same 
effect for gauge couplings. Consider for instance 
the covariant derivative of a $U(1)$, 
which is given by $D_M = \partial_M - i {\rm g} A_M$.
Since mass dimension of our gauge field is 
$[A_M] =1 + \frac{n}{2} $, thus gauge coupling has to have mass dimension  
$[{\rm g}] = - \frac{n}{2}$. 
We can explicitely write down the mass scale 
and introduce a new dimensionless coupling, $g_\ast$, as
 \be 
 {\rm g} = \frac{g_\ast}{M_\ast^{n/2}}
 \label{gast}
 \ee
To identify the effective coupling we just have to look at the zero mode level.
Consider for instance the gauge couplings of some fermion $\Psi$ 
for which the effective Lagrangian comes as
 \be 
 \int\!d^n y\, i\,\bar\Psi \Gamma^MD_M\Psi 
   = -i\, \frac{g_\ast}{\sqrt{M_\ast^n V_n}}
   A_\mu^{(0)}\bar\Psi_0 \Gamma^\mu\Psi_0 + \cdots~,
 \ee
where on the RHS we have used the generic property that the zero mode always 
comes with a volume suppression, $\Psi = \Psi_0/\sqrt{V_n} + \cdots$.  
Thus, if the effective coupling 
$g_{eff} = g_\ast/\sqrt{M_\ast^n V_n}$ is of order one, we are led to the
conclusion that $g_\ast$ must be at least as large as  $\sqrt{M_\ast^n V_n}$.

We must stress that this effective theories are
essentially non renormalizable for the infinite number of fields that they 
involve. However, the truncated theory that only considers a finite number
of KK modes is renormalizable. The cut-off for the inclusion of the
excited modes will be again the scale $M_\ast$. 

Non abelian bulk gauge theories follow  similar lines, 
although they involve the extra well known  ingredient of having 
interaction among gauge fields, which now will have a KK equivalent, 
where  vector  lines can either be zero or excited
modes, only restricted by conservation of the extra momentum (when it is
required) at each vertex.

\subsection{Short and large extra dimensions} 

SM fields may also reside on the extra dimensions, however, there is no
experimental evidence on the colliders of any KK excitation of any known
particle, that is well up to some hundred GeV. If SM particles are zero modes of
a higher dimensional theory, as it would be the case in string theory, the mass
of the very first excited states has to be larger than the current collider
energies. According to our above discussion, this would mean that the size of
the compact dimensions where SM particle propagate has to be rather small. 
This does not mean, however, that the fundamental scale has to be large.
A low $M_\ast$ is possible if there are at least two different classes of
compact extra dimensions: (i) short, where SM matter fields  can propagate, of
size $r$; and (ii) large of size $R$, tested by gravity and perhaps SM singlets. 
One can imagine the scenario as one where SM fields live in a 
$(3+\delta)$D brane, with $\delta$ compact dimensions, embedded in a larger
bulk  of $4+\delta+n$  dimensions.
In such a scenario, the volume of the compact space is given by 
$V_n = r^\delta R^n$, and thus, one can write the effective Planck scale as 
 \be
  M_P^2 = M_\ast^{2+\delta+n} r^\delta R^n~.
 \ee
Keeping $M_\ast$ around few tenths of TeV, only requires that the larger 
compactification scale $M_c=1/r$ be also about TeV,
provided $R$ is large enough, say in the submillimeter range. This way, 
short distance gravity experiments and collider physics could be complementary
to test the profile and topology of the compact space.

A priori, due to the  way the models had been constructed, 
there is no reason to belive that all SM particles could propagate in
the whole $4+\delta$ dimensions, and there are many different scenarios
considering a diversity of  possibilities. When only some fields do propagate
in the compact space one is force to first promote the gauge fields to the
bulk, since otherwise gauge conservation would be compromised. 
When all SM fields do feel such dimensions, the
scenario is usually referred as having Universal Extra Dimensions (UED). 

Phenomenology would of course be model dependent, and rather than making an
exhaustive review, we will just mention some general ideas. 
Once more, the effects of an extra dimensional nature of the fields can be 
studied either on the direct production of KK excitations, or through the
exchange of these modes in collider processes~\cite{coll2,review1,sm,quiros}. 
In non universal extra dimension
models KK number is not conserved, thus single 
KK modes can be produced directly in high energy particle collisions. 
Future colliders may be able to observe  resonances due to  KK modes  if
the compactification scale $1/r$  turns out to be on the TeV range. 
This needs a collider energy $\sqrt{s}\gs M_c = 1/r$. 
In hadron colliders (TEVATRON, LHC) 
the KK excitations might be directly produced in Drell-Yang processes 
$pp(p\bar p) \rightarrow \ell^-\ell^+X$ where the lepton pairs 
($\ell=e,\mu,\tau$) are produced via the subprocess $q\bar q\rightarrow
\ell^+\ell^+X$.  This is the more useful mode to search for 
$Z^{(n)}/\gamma^{(n)}$ even $W^{(n)}$. 
Current search for $Z'$ on this channels (CDF)
impose $M_c> 510~GeV$. Future bounds could be raised up to $650~GeV$ in
TEVATRON and $4.5~TeV$ in LHC, which with  100 $fb^{-1}$ of luminosity can
discover modes up to $M_c\approx 6~TeV$.

In UED models, due to KK number conservation, things may be more subtle 
since pair production of KK excitations would require more energy to reach the
threshold. 
On the other hand, the lighter KK modes would be stable and thus of easy
identification, either as large missing energy, when neutral, or as a heavy 
stable  particles if charged. It can also be a candidate for dark
matter~\cite{kkdm}

Precision test may be the very first place to look for constraints to the 
compactification scale $M_c$~\cite{coll2,sm,review1}. 
For instance, in non UED models, KK exchange
contributions to muon decay gives the correction to Fermi constant (see first
reference in~\cite{coll2})
\be
\frac{G_F^{eff}}{\sqrt{2}} = \frac{G_F^{SM}}{\sqrt{2}}
\left[1+ \frac{\pi^2}{3}~{m_W^2 r^2} \right]~;
\ee
which implies that $M_c\gs 1.6~TeV$.
Deviations on the cross sections due to virtual exchange of KK modes may
be observed in both, hadron and lepton colliders.  With a  $20~fb^{-1}$
of luminosity, 
TEVATRONII may observe signals up to $M_c \approx
1.3~TeV$. LEPII with a maximal luminosity of 200 $fb^{-1}$ could impose
the bound at $1.9~TeV$, while NLC may go up to 13 $TeV$, which slightly
improves the bounds coming from precision test.

{\it SUSY}.- 
Another  ingredient that may be reinstalled on the theory is supersymmetry.
Although it is not necessary to be considered for low scale gravity models, 
it is an interesting extension. 
After all, it seems plausible to exist  if the high energy
theory would be  string theory. If the theory  is supersymmetric, the effective
number of 4D supersymmetries increases due to the increment in the 
 number of fermionic degrees of freedom~\cite{sohnius}. 
For instance, in 5D  bulk fields come in $N=2$ 
supermultiplets~\cite{quiros,mirabelli}. The
on-shell field content of the a gauge supermultiplet is
${V}=(V_\mu,V_5,\lambda^i, \Sigma)$ where $\lambda^i\ (i=1,2)$ is a
symplectic Majorana spinor  and $\Sigma$ a real scalar in the adjoint
representation; $(V_\mu,\lambda^1)$ is  even under $Z_2$ and 
$(V_5,\Sigma,\lambda^2)$  is odd.  Matter  fields, on the other hand,
 are arranged in $N=2$ hypermultiplets that consist of  chiral and antichiral
$N=1$ supermultiplets. The chiral $N=1$ supermultiplets are even under   $Z_2$ 
and contain  massless states. These will correspond to  the SM fermions and
Higgses. 

Supersymmetry must be broken by some mechanisms that  gives masses to all
superpartners which we may assume are  of order $M_c$~\cite{quiros}. 
For some possible mechanism see Ref.~\cite{mirabelli}. 
In contrast with the case of four dimensional susy, where no extra effects
appear at tree level after integrating out the superpartners, in the
present case integrating out the  scalar field $\Sigma$ may induces a
tree-level contribution to  $M_W$~\cite{sm}, that could in principle 
be constraint by precision tests.

\subsection{Power Law Running of Gauge Couplings}

Once we have assumed a low  fundamental scale for quantum gravity, the
natural question is whether the former picture of a Grand Unified
Theory~\cite{guts} should be abandoned and with it a possible gauge theory
understanding of the quark lepton symmetry  and gauge hierarchy. On the
other hand, if string theory were the right theory above $M_\ast$  an unique
fundamental coupling constant would be expect, while the SM contains three
gauge coupling constants. Then, it
seems clear that, in any case, a sort of low energy  gauge coupling
unification is required. As pointed out in Ref.~\cite{ddg} and later
explored in~\cite{kaku,uni,uni2,unimore,kubo},  if the SM particles live
in higher dimensions such a low GUT scale could be realized.

For comparison let us mention how one leads to gauge unification in four
dimensions. Key ingredient in our discussion are the renormalization
group  equations (RGE) for the gauge coupling parameters that at one loop,
in the $\overline{MS}$ scheme, read
\be
\frac{d\alpha_i}{dt}=\frac{1}{2\pi} b_i \alpha^2_i
\label{rge}
\ee
where $t=ln \mu$. $\alpha_i = g_i^2/4\pi$; $i=1,2,3$, are the coupling
constants of the  SM factor groups $U(1)_Y$, $SU(2)_L$ and $SU(3)_c$
respectively.  The coefficient $b_i$ receives contributions from the 
gauge part and the matter including Higgs field  and its  completely
determined by 
 $4\pi b_i = {11\over 3} C_i(vectors) - {2\over 3}C_i(fermions)
 -{1\over 3}C_i(scalars)$,
where $C_i(\cdots)$ is the index of the representation to which the
$(\cdots)$ particles are assigned, and where we are considering Weyl
fermion and complex scalar fields. Fixing the normalization of the $U(1)$
generator as in the $SU(5)$ model, we get for the SM  $(b_1,b_2,b_3)
=(41/10,-19/6,-7)$ and for the Minimal Supersymmetric SM (MSSM)
$(33/5,1,-3)$.  Using Eq. (\ref{rge}) to extrapolate the values measured
at the $M_Z$ scale~\cite{pdg}: $\alpha^{-1}_1(M_Z)=58.97\pm .05$;
$\alpha^{-1}_2(M_Z)=29.61\pm .05$; and  $\alpha^{-1}_3(M_Z)=8.47\pm .22$,
(where we have taken for the strong coupling constant the global average),
one finds that only in the MSSM  the three couplings merge together at the
scale $M_{GUT}\sim  10^{16}$ GeV. This high scale naturally explains the
long live of the proton and  
in the minimal $SO(10)$ framework one gets  very
compelling  scenarios.

A different possibility for unification   that does not involve
supersymmetry is the existence of an intermediate left-right
model~\cite{guts} that breaks down to the SM symmetry at $10^{11-13}~GeV$.
It is worth mentioning that a non canonical
normalization of the gauge coupling may, however, substantially change
above pictures, predicting a different unification scale.
Such a different normalization may arise either in some no minimal (semi simple)
unified models, or in string theories where the SM group
factors are realized on non trivial Kac-Moody levels~\cite{ponce,kdienes}. 
Such  scenarios are in general
more complicated than the minimal $SU(5)$ or $SO(10)$ models since they
introduce new exotic particles. 

It is clear that the presence of KK
excitations will affect the evolution of couplings in gauge theories and
may alter the whole picture of unification of couplings. 
This question was first studied  by Dienes, Dudas and
Gherghetta (DDG)\cite{ddg} on the base of the effective
theory approach at one loop level. They found that 
above the compactification scale  $M_c$ one gets
 \be
 \alpha_i^{-1}(M_c) = \alpha_i^{-1}(\Lambda) + {b_i - \tilde{b}_i\over
2\pi}
 \ln\left( {\Lambda\over M_c}\right) + {\tilde{b}_i\over 4\pi}\
 \int_{w\Lambda^{-2}}^{wM_c^{-2}}\! {dt\over t}
 \left\{ \vartheta_3\left({it\over \pi R^2}\right)\right\}^\delta ,
 \label{exact}
 \ee
with $\Lambda$ as the ultraviolet cut-off and  
$\delta$  the number of extra
dimensions.
The Jacobi theta function
 $\vartheta(\tau) = \sum_{-\infty}^{\infty} e^{i\pi \tau n^2}$
reflects the sum over the complete tower. Here
$b_i$ are the beta functions of the theory below $M_c$,  
and $\tilde{b}_i$ are the contribution to the beta functions 
of the KK states at each excitation
level. 
The numerical factor $w$ depends on the renormalization scheme.
For practical purposes, we may approximate the above result by 
decoupling all the
excited states with masses above $\Lambda$, and assuming that the
number of KK states below certain energy $\mu$ between $M_c$ and
$\Lambda$ is well approximated by the volume of a $\delta$-dimensional
sphere of radius ${\mu\over M_c}$ given by
 $N (\mu,M_c) = X_\delta
 \left({\mu\over M_c}\right)^\delta$; 
with $X_\delta = \pi^{\delta /2}/\Gamma(1 +\delta /2)$.
The result is a power law behavior of the gauge coupling
constants~\cite{taylor}:
 \be
 \alpha_i^{-1}(\mu) = \alpha_i^{-1}(M_c) - {b_i - \tilde{b}_i\over 2\pi}
 \ln\left( {\mu\over M_c}\right) - {\tilde{b}_i\over 2\pi}\cdot
 {X_\delta\over\delta}\left[ \left({\mu\over M_c}\right)^\delta - 1\right],
 \label{ddgpl}
 \ee 
which  accelerates the meeting of the $\alpha_i$'s.
In the MSSM the energy range between $M_c$ and $\Lambda$ --identified as
the unification (string) scale $M_\ast$-- is relatively small due to the steep
behavior in the evolution of the couplings~\cite{ddg,uni}. For instance,
for a single extra dimension the ratio $\Lambda/M_c$ has an upper limit of
the order of 30, and it substantially decreases for larger $\delta$. 
This would, on the other hand, requires the short extra dimension where SM
propagates to be rather closer to the fundamental length.

This same relation can be understood on the basis of a 
step by step approximation~\cite{uni2}  as
follows. We take the SM gauge couplings and extrapolate their values up to
$M_c$, then, we add to the beta functions the contribution of the first KK
levels and  run the couplings upwards up to
just below the next consecutive level where
we stop and add the next KK contributions, and so on, 
until the energy $\mu$.
Despite the complexity of the spectra, the degeneracy of each level 
is always computable and performing a level by level approach of the gauge
coupling running is  possible.
Above the  $N$-th level the running  receives
contributions from $b_i$ and of
all the KK excited states in the
levels below, in total $f_\delta(N) = \sum_{n=1}^N g_\delta(n)$,
where $g_\delta(n)$ represent the total degeneracy of the level $n$. 
Running for all the first $N$ levels  leads to
 \be
 \alpha_i^{-1}(\mu) = \alpha_i^{-1}(M_c) - {b_i\over 2\pi}
 \ln\left( {\mu\over M_c}\right) - {\tilde{b}_i\over 2\pi} \left[
 f_\delta(N)\ln\left(\mu\over M_c\right) - 
 {1\over 2}\sum_{n=1}^N g_\delta(n)\ln n  \right].
 \label{hdlog}
 \ee
A numerical comparison of this expression with the
power law running 
shows the accuracy of that approximation. Indeed, in the
continuous limit the last relation  reduces into Eq. (\ref{ddgpl}).
Thus, gauge coupling unification may now  happen at TeV
scales~\cite{ddg}.

Next, we will discuss how accurate this unification is.
Many features of unification can be studied
without bothering about the detailed subtleties of the running. 
Consider the generic form for the 
evolution equation 
 \be
 \alpha_i^{-1}(M_Z) = \alpha^{-1} + {b_i\over 2\pi}
 \ln \left( {M_\ast\over M_Z}\right) + {\tilde{b}_i\over 2\pi} 
 F_\delta\left( {M_\ast\over M_c}\right), 
 \label{rgef}
 \ee 
where we have changed $\Lambda$ to $M_\ast$ to keep our former notation.  
Above, $\alpha$ is the unified coupling and $F_\delta$ is  given by
the expression between parenthesis in Eq. (\ref{hdlog}) or its
correspondent limit in Eq. (\ref{ddgpl}).
Note that the information that comes from the bulk is being
separated into two independent parts: all the structure of the KK spectra 
$M_c$ and $M_\ast$ are  completely embedded into the
$F_\delta$ function, and  their contribution is actually
model independent. The only (gauge)  model dependence comes in the beta
functions, $\tilde{b}_i$. 
Indeed, Eq. (\ref{rgef}) is similar to that of  the two step unification
model where a new gauge symmetry appears at an intermediate energy scale.
Such models are very constrained by the one step unification  in the MSSM.
The argument goes as follows: let us define the vectors: 
${\bf b}= (b_1,b_2,b_3)$; 
$\tilde{\bf b}= (\tilde{b}_1,\tilde{b}_2,\tilde{b}_3)$; 
${\bf a} = (\alpha_1^{-1}(M_Z),\alpha_2^{-1}(M_Z),\alpha_3^{-1}(M_Z))$ and 
${\bf u} = (1,1,1)$, and construct the unification barometer~\cite{uni2}
$\Delta\alpha\equiv ({\bf u}\times {\bf b})\cdot {\bf a}$.
For single step unification models the unification
condition amounts to have $\Delta\alpha=0$.
As a matter of  fact, 
for the SM $\Delta\alpha=41.13\pm 0.655$, while for the MSSM
$\Delta\alpha = 0.928\pm 0.517$, leading to unification within two
standard deviations. In this notation 
Eq. (\ref{rgef}) leads to 
 \be
 \Delta\alpha = 
 [({\bf u}\times {\bf b})\cdot \tilde{\bf b}]\ {1\over 2 \pi}F_\delta.
 \label{dalpha}
 \ee
 Therefore, for the MSSM, 
we get the constrain~\cite{mohapatra}
 \be
 (7\tilde{b}_3 - 12\tilde{b}_2 + 5 \tilde{b}_1)F_\delta= 0.
 \ee
There are two solutions to the this equation: 
(a) $F_\delta(M_\ast/M_c)=0$, which means $M_\ast = M_c$,
bringing  us back to the MSSM by pushing up the compactification scale to
the unification scale.
(b) Assume that the beta coefficients $\tilde{b}$ conspire to 
eliminate
the term between brackets: 
$(7\tilde{b}_3 - 12\tilde{b}_2 + 5 \tilde{b}_1) = 0$,
or equivalently~\cite{ddg}
 \be
 {B_{12}\over B_{13} } = {B_{13}\over B_{23} } = 1; \qquad \mbox{where}\qquad
 B_{ij} = {\tilde{b}_i- \tilde{b}_j\over b_i- b_j} .
 \ee
The immediate consequence of last possibility is the indeterminacy  of
$F_\delta$, which means that we may put $M_c$ as a free parameter in the
theory. For instance we could choose $M_c\sim$ 10 TeV to maximize the
phenomenological impact of such models.  It is compelling to stress that
this conclusion is independent of the explicit form of $F_\delta$.
Nevertheless, the minimal model where all the  MSSM particles propagate on
the bulk does not satisfy that constrain~\cite{ddg,uni}. Indeed, in this
case we have $(7\tilde{b}_3 - 12\tilde{b}_2 + 5 \tilde{b}_1) = -3$, which
implies a  higher prediction for $\alpha_s$ at low $M_c$. As lower the
compactification scale, as higher the prediction for $\alpha_s$. 
However,  as
discussed  in Ref.~\cite{uni} there are some  scenarios where the
MSSM fields are distributed in a nontrivial  way among the bulk and the
boundaries which lead to unification. There is also the  obvious
possibility of adding  matter to the MSSM to correct the accuracy on
$\alpha_s$. 

The SM case has similar complications. 
Now Eq. (\ref{rgef}) turns out to be a system of three equation with three
variables, then, within the experimental accuracy on $\alpha_i$, 
specific predictions for $M_\ast$, $M_c$ and $\alpha$ will arise. 
As $\Delta\alpha\neq 0$, the
above constrain does not apply, instead the matter content should
satisfy the consistency conditions~\cite{uni2}
 \be 
 Sign(\Delta\alpha)  
 = Sign[({\bf u}\times {\bf b})\cdot \tilde{\bf b}]
 = - Sign(\tilde{\Delta}\alpha) \ ;
 \label{cons2}
 \ee
where $\tilde{\Delta}\alpha\equiv({\bf u}\times\tilde{\bf b})\cdot{\bf a}$.
However, in the minimal model where all SM fields are assumed to have KK
excitations  one gets 
$\tilde{\Delta}\alpha = 38.973\pm 0.625$; and 
$({\bf u}\times {\bf b})\cdot \tilde{\bf b}^{SM} = 1/15$. 
Hence, the constraint (\ref{cons2}) is not
fulfilled and unification does not occur. 
Extra matter could of course improve this 
situation~\cite{ddg,uni,uni2}. Models with non canonical normalization may
also modify this conclusion~\cite{uni2}. A particularly
interesting outcome in this case is that
there are some cases where, without introducing extra matter at the SM
level, the unification scale comes out to be
around $10^{11}$ GeV (for instance $SU(5)\times SU(5)$,
$[SU(3)]^4$ and $[SU(6)]^4$). 
These models fit nicely into the  intermediate string scale models
proposed in~\cite{quevedo}, and also with the expected scale in
models with local $B-L$ symmetry. 
High order corrections has been considered in Ref.~\cite{unimore}. 
It is still possible that this joint to 
threshold corrections  might  correct the situation and  improve  
unification, so one can not rule it out on the simple basis of one-loop running.  
Examples of he analysis for
the  running of other coupling constants could be found
in~\cite{ddg,kubo}. Two step models were also studied in~\cite{uni2}.

It has been argue recently by some authors 
that precision power-law
unification can not be
claimed in this context because of the strong UV
sensitivity of the 5D $F^2$ terms of the three SM factor groups 
(see for example Refs.~\cite{pilo} and references therein). 
As a result no precise calculation of the ratio of low energy gauge couplings 
could be made without a UV completition of the higher dimensional theory
(see also Refs~\cite{seiberg}).
Some details and more references on this 
are also found in Refs.~\cite{hebecker}, 
where it has also been 
pointed out a class of models where 
power-law unification can work in a quantitatively controlled way 
(the main idea being soft GUT breaking in 5D field theory). 
Extra information beyond the
4D low-energy particle spectrum has, in any case, to be supplied. 
The issue of power law running was also discussed in connection with
deconstructed theories and warped 5D models in Refs.~\cite{chankowski,rothstein}

\section{Symmetry Breaking with Extra Dimensions}

Old and new ideas on symmetry breaking 
have been revisited  and further developed in the 
context of  extra dimension models by many authors in the last few years. 
Here we provide a short overview of this topic. Special attention is payed to 
spontaneous symmetry breaking, 
and the possible role compactification may play to induce the 
breaking of some continuous symmetries. An extended review can be found in 
the lectures by M. Quiros in reference~\cite{reviews}.

\subsection{Spontaneous Breaking}
 
The simplest place to start is reviewing the  spontaneous symmetry breaking
mechanism  as implemented with  bulk fields, as it would be the case in a higher
dimensional SM. Let us consider the usual potential for a bulk scalar field
 \be  
 V(\phi) = -\frac{1}{2}m^2 \phi^2 + 
 \frac{\lambda_\ast}{4{M_\ast}^n}\phi^4~.
 \ee
First thing to notice is the suppression on the  quartic coupling. Minimization
of the potential gives the condition
 \be 
   \la\phi\ra^2 = \frac{m^2 {M_\ast}^n}{\lambda}~.
 \ee
RHS of this equation is clearly a constant, which means that in the absolute
minimum  only the zero mode is picking up a vacuum expectation value (vev). 
As the mass parameter $m$ is naturally smaller than the fundamental scale
$M_\ast$, this naively implies that the minimum has an enhancement respect 
to the standard 4D result. Indeed, if one considers the KK expansion 
$\phi = \phi_0/\sqrt{V_n} + \cdots$; is easy to see that the effective
vacuum as seen in four dimensions is
 \be 
    \la\phi_0\ra^2 = \frac{m^2 {V_n M_\ast}^n}{\lambda} = 
    \frac{m^2}{\lambda_{eff}}~.
 \ee
The enhancement can also be seen as due to the suppression of the effective
$\lambda_{eff}$ coupling.
The result can of course be verified if calculated directly in the effective
theory (at zero mode level) obtained after integrating out the extra dimensions.
Higgs mechanism, on the other hand, happens as usual. 
Consider for instance a bulk $U(1)$ gauge, 
broken by the same scalar field we have 
just discussed above. The relevant terms contained in the kinetic terms,
$({\cal D}_M\phi)^2$, are as usual ${\rm g}^2 \phi^2 A_M(x,y) A^M(x,y)$.
Setting in the  vev and Eq.~(\ref{gast}) one gets the global mass term 
 \be 
 \frac{g_\ast^2}{V_nM_\ast^n}\la\phi_0\ra^2A_M(x,y) A^M(x,y) 
 = g_{eff}^2\la\phi_0\ra^2A_M(x,y) A^M(x,y)~. 
 \ee
Thus, all KK modes of the gauge field acquire a 
universal mass contribution from the bulk vacuum.

\subsection{Shinning vevs}  

Symmetries can also be broken at distant branes, 
and the breaking be communicated by the mediation of bulk fields to some other
brane~\cite{break}. 
Consider for instance the following toy model. We take a brane located
somewhere in the bulk, let say at the position $\vec{y}=\vec{y}_0$, where 
there is a localized scalar field, $\varphi$, which couples to a bulk scalar
$\chi$, such that the Lagrangian in the complete theory is written as
 \be 
 {\cal L}_{4D}(\varphi)\delta^n(\vec y-\vec y_0) 
 + \frac{1}{2}\partial_M \chi\partial^M \chi 
 - \frac{1}{2}m_\chi^2\chi^2 - V(\phi,\chi)~.
 \ee
For the brane field we will assume the usual Higgs potential 
$V(\varphi) = -\frac{1}{2}m_\varphi^2 \varphi^2 + \frac{\lambda}{4}\varphi^4$, 
such that $\varphi$ gets a non zero vev. For the interaction potential we take 
 \be 
 V(\varphi,\chi) = \frac{\mu^2}{M_\ast^{n/2}}\, \varphi\,\chi\,
 \delta^n(\vec{y}-\vec y_0)~;
 \ee
with $\mu$ a mass parameter. Thus, 
 $\la\varphi\ra$ acts as a point-like source for $\la\chi\ra(\vec y)$,
 \be 
 \left(\nabla_\bot^2 + m_\chi^2\right)\la\chi\ra(\vec y) = 
-\frac{\mu^2}{M_\ast^{n/2}}\la\varphi\ra\, \delta^n(\vec{y}- \vec y_0)~.
 \ee
The equation has the solution
 \be  
 \la\chi\ra(\vec y) = \Delta(m_\chi;\vec y_0- \vec y)\la\varphi\ra~,
 \ee
with $\Delta(m_\chi;\vec y_0- \vec y)$ the physical propagator of the field.
Next we would be interested in what a second brane localized in $\vec{y}=0$
would see, for which we introduce the coupling of the bulk field to some fermion
on the second brane, 
$\frac{h}{\sqrt{M_\ast^n}}\bar\psi^c\psi\, \chi\, \delta^n(\vec{y})$. 
If we assume that all those fields carry a global $U(1)$ charge, this last
coupling will induce the breaking of such a global symmetry on the second brane,
by generating a mass term
\be 
\left[\frac{\Delta(m_\chi;\vec y_0)}{\sqrt{M_\ast^n}}\right]\, 
h~\la\varphi\ra\bar\psi^c\psi
\ee
Since this requires
the physical propagation of the information through the distance, the second
brane sees a suppressed effect, which results in a small breaking of
the $U(1)$ symmetry. 
This way, we get a suppressed effect with out the use of large energy scales. 
This idea has been used where small vevs are needed, as for instance to produce 
small neutrino masses~\cite{shinning,ma}. It may also be used to induce small
SUSY breaking terms on our brane~\cite{susyb}.

 \subsection{Orbifold Breaking of Symmetries}

We have mentioned in previous sections that by orbifolding the extra dimensions 
one can get chiral theories. In fact, orbifolding can actually do more than
that. It certainly projects out part of the degrees 
of freedom of the bulk fields via the imposition of the extra discrete
symmetries that are used in the construction of the orbifold out of the compact
space. However it gives enough freedom as to choose which components of the
bulk fields are to remain  at zero mode level. In the case of fermions on 5D,
for instance, we have already commented that under 
$Z_2$ the fermion generically transform as $\Psi \rightarrow \pm \gamma_5 \psi$,
where the $\pm$ sign can be freely chosen. The complete 5D theory is indeed
vector-like since chirality can not be defined, which  means the theory is
explicitely left-right symmetric. Nevertheless, when we look up on the zero mode
level, the theory would have 
less symmetry than the whole higher dimensional theory, since only a left (or
right) fermion do appears.
This can naturally be used to break both global and 
local symmetries~\cite{parity,orbi} and so, it has been extendedly  exploited   
in model building. Breaking of parity due to the projection of part of the
fermion components with well defined 4D chirality is just one of many examples.
A nice model where parity is broken using bulk scalars was presented in
Ref~\cite{parity}, for instance. In what follows we will consider the case of
breaking  non abelian symmetries through a simple example.

{\it  Toy Model: Breaking $SU(2)$ on $U(1)/Z_2$}.-
Consider the following simple  5D model. We take a bulk scalar doublet 
 \be 
 \Phi= \left(\begin{array}{c} \phi \\ \chi \end{array}\right)~,
 \ee
and assume for the moment that the $SU(2)$ symmetry associated to it is  global. 
Next, we assume the fifth dimension is compactified on the
orbifold $U(1)/Z_2$, where, as usual, $Z_2$ means the identification of points
$y\rightarrow -y$. For simplicity we  use $y$ in the unitary circle defined
by the interval $[-\pi,\pi]$ before orbifolding. 
$Z_2$ has to be a symmetry of the  Lagrangian, and that
is the only constrain in the way $\Phi$ should transforms under $Z_2$.
As the scalar part of the 
Lagrangian   goes as ${\cal L} = \frac{1}{2}({\partial}_M\Phi)^2$, the 
most general transformation rule would be
 \be
 \Phi \rightarrow  P_g\Phi~;
 \label{pg}
 \ee
where $P_g$ satisfies $P_g^\dagger=P_g^{-1}$, thus, the simplest 
choices are  $ P_g = \pm {\bf 1};~ \pm \sigma_3$  up to a global phase, that we
will neglect for the moment. Clearly the first option only means taking both
fields on the doublet to be simultaneously even or odd, with no further
implication for the theory. However, the second choice is some what more
interesting. Taking  for instance $P_g = \sigma_3$, this selection explicitely
means that under $Z_2$
 \be 
 \left(\begin{array}{c} \phi \\ {\chi} \end{array}\right)
 \rightarrow \left(\begin{array}{c} \phi \\ { -\chi} \end{array}\right)~.
 \label{orbi1}
 \ee
That is, $\chi$ is force to be an odd field . Therefore, 
at the zero mode level, one would only see the 
$\phi$ field, and thus, the original $SU(2)$ symmetry would not be  evident. 
In fact, the lack of the whole symmetry would be clear by looking at the whole
KK spectrum, where at each level there is not appropriate pairing of fields 
that may form a doublet. Either $\phi_n$ or $\chi_n$ is missing. 

The effect of this non trivial  $Z_2$ transformation 
can also be understood via the boundary conditions. Whereas $\phi$ has been
chosen  to be an even field, whose KK expansion only contains cosine functions
which are non zero at both  ends of the  space, located at $y = 0$ and $y=\pi$; 
$\chi$ vanishes at those points, $\chi(0) = \chi(\pi) =0$. Hence the boundaries
are forced to have  less symmetry than the bulk, in fact only a residual $U(1)$
symmetry,  which  is reflected in the effective theory. Thus, the selection of
the orbifold condition (\ref{orbi1})  results in the effective 
breaking of $SU(2)$ down to $U(1)$. 
We should notice that the transformation (\ref{pg}) is  an
inner automorphism which triggers a  breaking that  preserves the range 
of the original group. 
If fact all inner automorphism do. To reduce the rank of the group one can use
outer automorphism  (see for instance Hebecker and  March-Russell in
Ref.~\cite{orbi} and Quir\'os in Ref.~\cite{reviews}).

Let us now see what happens if $SU(2)$ is assumed to be a local symmetry.
In this case we should ask the covariant derivative ${\cal{D}}_M{\Phi}$ 
to have proper $Z_2$ transformation rules: 
\be {\cal{D}}_\mu{\Phi_a} \rightarrow P_g {\cal{D}}_\mu \Phi_a
\qquad\mbox{and}\qquad 
{ {\cal{D}}_5{\Phi_a} \rightarrow {-P_g}{\cal{D}}_5 \Phi_a} ~.
\ee
This fixes the way the gauge fields,
${\cal{A}}_M = A_M^a ~{\sigma}^a/\sqrt{2}$,
should transform:
 \be  
 {\cal{A}}_\mu \rightarrow P_g{\cal{A}}_\mu P_g^{-1} 
 \qquad\mbox{and}\qquad 
 {\cal{A}}_5 \rightarrow -P_g {\cal{A}}_5 P_g^{-1}
 \ee
Now, for $P_g=\sigma_3$ we get the following assignment of  parities:
$ W_{\mu}^3 (+)$; $W_{\mu}^\pm (-)$; $W_5^3 (-)$ and $W_5^\pm (+)$.
Clearly, as only even modes are non zero at the boundaries, 
only the $U(1)$ associated to $W_{\mu}^3$ remains intact, as expected. 

Notice that  we are projecting out the zero mode of the 
charged vector bosons to the price of leaving instead two massless
charged scalars $W_5^\pm$ in the effective theory. These extra fields can be
removed by a further orbifolding of the compact space. Indeed if one uses
instead the orbifold $U(1)/Z_2\times Z'_2$, where the second identification of
points is defined by the transformation $Z'_2: y'\rightarrow -y'$, where 
$y'= y+\frac{\pi}{2}$, one can freely choose another set of parities for the 
field components in the doublet, corresponding to the transformation properties
of the doublet under $Z'_2: \Phi\rightarrow P'_g \Phi$. Therefore, the KK 
wave functions along the fifth dimensions will be now  classified  under both
these parities. We will then have, 
\be 
\ba{ccc}
 \xi^{(+,+)} \sim\cos(2n\,y/R) &\qquad \xi^{(+,-)} \sim\cos[(2n-1)\,y/R]\\[1ex]
\xi^{(-,+)} \sim\sin(2n\,y/R) &\qquad \xi^{(-,-)} \sim\sin[(2n-1)\,y/R]\ea
\ee
up to a normalization factor. Clearly, only the completely even function,
$\xi^{(+,+)}$, do contain a zero mode. 
If we now take  
the transformations to be given by $P_g= {\bf 1}$ and  $P'_g= \sigma_3$ for 
$Z_2$ and $Z'_2$ respectively, then, we then get the parity assignments
$W^3_\mu(+,+)$;  $W^\pm_\mu(-,+)$; but 
$W^3_5(-,-)$  and  $W^\pm_5(+,-)$. Therefore, at zero mode level, only
$W^3_\mu$ would appear.

 \subsection{Scherk-Schwarz mechanism}

When  compactifying, one assumes that the extra dimensions form a
quoting space $C=M/G$, which is constructed out of a non compact manifold
$M$  and   a discrete group $G$  acting on $M$, with the identification of
points $ y\equiv \tau_g(y)$ for $\tau_g$ a representation of $G$, which means
that $\tau_{g_1}\tau_{g_2} = \tau_{g_1}\cdot\tau_{g_2}$. $G$ should be acting
freely, meaning that only $\tau_e$ has fixed points in $M$, where $e$ is 
the   identity in $G$. $M$ becomes the covering space of $C$.
After the identification physics should not depend on individual
points in $M$ but only on on points in $C$ (the orbits), such that 
${\cal L} [\phi(y)] = {\cal L} [\phi(\tau_g(y))]$.
To satisfy this, in ordinary compactification one 
uses the  sufficient condition  $\phi(y) = \phi(\tau_g(y))$. 
For instance, if we use $\tau_n(y) = y+2n\pi$
for $y\in {\cal R}$ and $n$ an integer number, the identification leads to the
fundamental interval $(y,y + 2\pi]$ that is equivalent to the unitary circle. 
The open
interval only states that both the ends describe the same point. One  
usually writes a close interval with the implicit equivalence of ends.
Any choice of
$y$ leads to a equivalent fundamental domain in the covering space $\cal R$.
One can take for example $y=-\pi$ so that the intervale becomes $(-\pi,\pi]$.

There is, however, a more general
necessary and sufficient condition for the invariance of the Lagrangian 
under the action of $G$, which is given by the so called Scherk-Schwarz
compactification~\cite{ss}
 \be  
 \phi(\tau_g(y))=T_g\phi(y)~,
 \ee
where $T_g$ is a representation  of $G$ acting on field space, usually called
the {\it twist}. 
Unlike ordinary compactification, given for a trivial twist, for Scherk-Schwarz
compactification twisted fields are not single value functions on $C$.
$T$ must be an operator corresponding to a symmetry of the Lagrangian.
A simple example is the  use the $Z_2$ 
symmetry  for which the twisted condition would be  
$\phi(-\pi) \equiv T \phi(\pi) = -\phi(\pi)$.  

Notice that the orbifold is somewhat a step beyond compactification. 
For orbifolding we take a compact manifold $C$ and a discrete group 
$H$ represented by some operator $\zeta_h$ acting non freely on $C$.
Thus, we mode out $C$ by identifying points on $C$ such that 
$y\equiv \zeta_h(y)$,  for some $h$ on $H$, and require that fields 
defined at these two points differ by some transformation $Z_h$, 
$\phi(x,\zeta_h(y)) = Z_h \phi(x,y)$,
which is a symmetry of the theory. The resulting space $C/H$ is not a smooth
manifold but  it has singularities at the fixed points.

Scherk-Schwarz boundary conditions can change the properties of the 
effective 4D theory and can also be used to break some symmetries of the
Lagrangian. Consider the simple toy model where we take $M={\cal R}$, and 
the group $G={\cal Z}$, as for the circle. Thus we use the identification 
on $\cal R$, $\tau_n(y) = y+2n\pi R$, with $R$ the radius of the circle 
as before. The group $\cal Z$ has an infinite
number of elements, but all of them can be obtained from just one generator, 
the simple translation by $2\pi$. Thus, there is  only one independent twist, 
$ \phi(y+2\pi R) = T\phi(y)$ and other elements of $\cal Z$ are just given
by $T_n = T^n$. We can then choose the transformation to be 
 \be 
 \phi(y+2\pi R) = e^{2\pi\, i\, \beta}\phi(y)
 \ee
where $\beta$ is called the Scherk-Schwarz charge.
Thus, with this transformation rule instead of the usual 
Fourier expansion for the fields we get
 \be 
 \phi(x,y) = 
 e^{i\, \beta y/R}\sum_{n= -\infty}^{\infty}\phi_n(x)\, e^{i\,n\,y/R}~.
 \ee
 
At the level of the KK excitations we see that fifth momentum
is less trivial than usual, indeed, acting  $p_5 = -i\partial_y$
on the field one  sees that the KK mass is now given as
 \be 
 m_n =\frac{n+\beta}{R}~.
 \ee
Therefore, in this model  all modes are massive, including the zero mode. 
This particular property can be used to break global  symmetries. 
For instance, if we assume a Global $SU(2)$, and consider a doublet
representation of fields 
 \be 
 \Phi= \left(\begin{array}{c} \phi \\ \chi \end{array}\right)~;
 \ee 
one can always choose the non trivial twist 
\be 
\Phi  \longrightarrow
\left(\begin{array}{c} { e^{i\, \beta y/R}}\phi \\ 
          \chi \end{array}\right)~;
\ee
which explicitely breaks the global symmetry. Whereas the zero mode
of $\chi$ appears massless, this does not happen for $\phi$.
Moreover,  the effective theory does not  present  the $SU(2)$ symmetry 
at any level.

Similar Scherk-Schwarz mechanism can be used to break local gauge 
symmetries. The result is actually equivalent to the so called Wilson/Hosotani
mechanism~\cite{hosotani,cheng} 
where the  $A^{3}_5$ component of the gauge vector, $A^3_M$, 
may by some dynamics acquire a non zero vev, 
and induce a mass term for the 4D  gauge fields, $A^a_\mu$, through 
the term 
${\la A^3_{5} \ra^2}\, (A^{\mu\,1}\, A^1_{\mu}+ A^{\mu\,2}\, A^2_{\mu})$, 
which is contained in $Tr\, F_{MN} F^{MN}$. Thus $SU(2)$ would be
broken down to $U(1)$.
Another  interesting use of this mechanism could be the breaking of
supersymmetry~\cite{delgado}.
For more discussions  see  M. Quir\'os in Ref.~\cite{reviews} and references
therein.

\section{$\nt{L}$ and $\nt{B}$ in Low Gravity Scale Models}
Baryon number ($B$) and Lepton number ($L$) are conserved
quantities in the SM.  However, it is believe that such global symmetries
may not be respected by the physics that lays beyond electroweak scale. 
Well known examples are GUT theories, which contain new quark-lepton
interactions that violate baryon number. R parity breaking terms in
supersymmetric theories usually include lepton and baryon number violation too. 
It is also believe that quantum gravity would not conserve any global
symmetry. 

Regarding lepton number, 
several experiments have provided conclusive evidence for the oscillation of
neutrinos, and this only  takes place if neutrinos 
are massive~\cite{nus}. The most appealing 
four dimensional mechanisms which generates masses for the SM 
left handed neutrino  is see-saw, which also introduce 
right handed Majorana masses that violate lepton number.
The generated mass appears  effectively
through the non renormalizable operators
 \be
   \frac{(LH)^2}{\Lambda}~, 
 \ee
where $L$ is the lepton doublet and $H$ the Higgs field.
 In order to get the right order for the mass one has to 
invoke high energy physics with scales
about $\Lambda\sim 10^{13}$ GeV or so. That should be
the mass scale for right handed neutrinos.

Possible evidence of the violation of baryon number can be found in the 
baryon domination in the universe. The simplest effective operator that would
produce proton decay, for instance, has the form
 \be 
 \frac{\bar Q^c Q \bar Q^c L}{\Lambda^2}~,
 \ee
where $Q$ stands for the quark representations and color index sum is implicit. 
Since the proton has a life time larger than $10^{33}$ yrs., 
(for the decay into a pion and a positron), this implies that the 
suppression on this operator has to be large enough, in fact 
$\Lambda\ls 10^{16}$ GeV.

Obviously, with a fundamental  scale at the TeV range, 
understanding the small neutrino masses and controlling proton decay 
poses a  theoretical challenge to the new  theories. 
The problem seems worsen because, given that the SM has to be treated only as an
effective theory, one is in principle entitled to write all operators that are
consistent with the known symmetries of the theory. 
However, because now  the non renormalizable
operators can only be  suppressed  by powers of $1/M_\ast$,  the effects
of this operators may be greatly enhanced.  Particularly, 
neutrino mass would be  large,
of  order $\la H \ra^2/M$, whereas proton decay may be too fast.  

To exclude or properly suppress 
these operators  one has to make additional assumptions, or work out 
explicit models. Unfortunately, it is not possible to elaborate here on all
ideas in the literature. Thus, we will rather just comment some  
interesting possibilities, 
given particular examples in the case of neutrino masses.  
For proton decay, on the other hand, we shall
discuss the two ideas we believe are the most
promising: 6D orbifolded theories~\cite{ponton} 
and wave function localization~\cite{sch1}.

\subsection{Neutrino Mass Models}

Regarding neutrino mass, the simplest way to control the unwanted operators 
is by adding lepton number as a real symmetry, whose eventual breaking should
generate only small masses. We can classify the models into  two classes 
depending on whether lepton number, or equivalently $B-L$, 
is a global or local symmetry. 

\subsubsection{Models with global $L$ symmetry.}

In the context of models that have  a global $U(1)_{L}$ symmetry, one
can  get small neutrino masses by introducing isosinglet neutrinos in the
bulk~\cite{dienes1} which carry lepton number.  As this is a sterile
neutrino,  it comes natural to assume that it may propagate into the bulk
as well as gravity, while the SM particles remain attached to the brane.
These models are interesting since they lead to small neutrino masses
without any extra assumptions. 

Let  $\nu_B( x^{\mu}, y)$ be a bulk neutrino, living on the $U(1)/Z_2$ orbifold, 
which we take to be massless 
since the Majorana mass violates conservation of Lepton number  and the
five dimensional Dirac mass is  forbidden by the orbifold symmetry. 
This neutrino couples  to the
standard lepton doublet and  to the  Higgs field  via 
${h\over \sqrt{M}}\bar{L} H \nu_{BR}~\delta(y)$. Once the Higgs develops its
vacuum, this coupling  generates 
the four dimensional Dirac mass terms
 \be
  m\bar\nu_L\left( \nu_{0R} + \sqrt{2}\sum_{n=1}^{\infty} \nu_{nR}\right) ,
 \label{l1}
 \ee  
where the mass $m$ is given by~\cite{dvali2}
 \be 
 m = h v {M_\ast\over M_{P}}\sim 10^{-2}~eV\times {hM_\ast\over 100~TeV}.
 \ee
Therefore, if $M_\ast\sim 100~ TeV$ 
we get just the right order of magnitude on the
mass as required by the experiments. Moreover,   even if 
the KK decouple for a small  $R$, we will still  get the
same Dirac mass for $\nu_L$ and $\nu_{0R}$, as far as $M_\ast$ remains in the
$TeV$ range. 
The general result is actually $R$ independent, provided the bulk neutrino
propagates in the whole bulk.
After including the KK masses, we may write down all
mass terms in the compact form~\cite{mnp}
 \be 
   (\bar{\nu}_{eL}  \bar{\nu}'_{BL})\left(\begin{array}{cc}
 m &\sqrt{2} m\\ 0 & \partial_5
 \end{array}\right)\left(\begin{array}{c}\nu_{0B} \\
 \nu'_{BR}\end{array}\right),
 \label{m1}
 \ee
where the notation is as follows:  $\nu'_B$ represents the KK
excitations.  The off diagonal term  $\sqrt{2} m$  is actually an infinite
row vector of the form $\sqrt{2} m (1,1,\cdots)$ and the operator
$\partial_5$ stands for the diagonal and infinite KK mass matrix whose
$n$-th entrance is  given by $n/R$.  
Using this short hand notation  it is straightforward  to
calculate the exact eigensystem for this mass matrix~\cite{mp2}. 
Simple algebra yields the characteristic equation
 $2 \lambda_n = \pi \xi^2 \cot(\pi \lambda_n)$,
with $\lambda_n=m_nR$, $\xi=\sqrt{2}mR$, and where $m_n$ is the mass 
eigenvalue~\cite{dienes1,dvali2}. 
The weak eigenstate is given in terms of the mass eigenstates, 
$\tilde \nu_{nL}$, as 
 \be
 \nu_L = \sum_{n=0}^\infty {1\over N_n} \tilde \nu_{nL} ,
 \label{nul}
 \ee
where the mixing $N_n$ is  given by
$N^2_n = \left(\lambda_n^2 + f(\xi)\right)/\xi^2$,
with $f(\xi)= \xi^2/2 + \pi^2 \xi^4/4$~\cite{mp2}.
Therefore, $\nu_L$ is actually  a coherent superposition of an
infinite number of massive modes. As they evolve
differently on time, the above equation will give rise to neutrino
oscillations, $\nu\rightarrow \nu_B$, 
even though there is only one single flavor. This is a
totally new effect that was thought it may be an alternative to explain neutrino
anomalies, unfortunately it does not seem to be detectable in
current neutrino experiments.  An analysis of the implications of the mixing
profile in these models  for solar neutrino deficit was presented
in~\cite{dvali2}.  Implications for atmospheric neutrinos were
discussed in~\cite{barbieri}, and some early 
phenomenological bounds were given
in~\cite{barbieri,pospel}. A comprehensive analysis for 
three flavors is given in~\cite{3nu}. 
Overall, 
the non observation of any  effects attainable to extra dimensional
oscillations means that the first excited level (so the tower)
is basically decoupled from the zero mode, which means that $1/R\gs 10 eV$, or
equivalently $R\ls 10^{-2}~\mu m$.  

The extension of this model to three
brane  generations, $\nu_{e,\mu,\tau}$, is straightforward.  
However, to give masses to the three standard generations
three bulk neutrinos are needed~\cite{mp2}. 
This comes out  from the fact that with a
single bulk neutrino only one massless right handed neutrino is present
(the zero mode), then, the coupling to brane fields will generate only one
new massive Dirac neutrino.  After introducing a rotation by 
an unitary matrix
$U$ on the weak sector, the most general Dirac  mass terms with three
flavors and arbitrary Yukawa couplings may be written down as 
 \be 
 -{\cal L} =
 \sum_{\alpha=1}^3\left[ 
  m_\alpha \bar\nu_{\alpha L} \nu^\alpha_{BR}(y=0) + 
  \int dy\, \bar\nu^\alpha_{BL}\partial_5 \nu^\alpha_{BR} + 
  h.c. \right] ,
 \label{three} 
 \ee
where
$\nu_{aL}=U_{a\alpha}\nu_{\alpha L}$, 
with $a= e,\mu,\tau$ and $\alpha=1,2,3$. 
The  mass parameters $m_\alpha$ are the
eigenvalues of the 
Yukawa couplings matrix multiplied by the vacuum $v$, and as stated before
are naturally of the order of eV or less.
This reduces the analysis to
considering three sets of mixings given as in the previous case. Each set
(tower) of mass eigenstates is characterized by
its own parameter $\xi_{\alpha}\equiv \sqrt{2}m_{\alpha} R$.
When this is small only the mass terms that involve zero mode would be relevant,
and one indeed gets three Dirac massive neutrinos, $\nu_{\alpha}$, with  
the mixing angles given by $U_{a\alpha}$.

\subsubsection{Models for Majorana masses.}

Some extended scenarios that consider the  generation of  Majorana masses
from the spontaneous breaking of lepton number either on the bulk or on 
a distant brane have been
considered in Refs.~\cite{ma,beta}. Shinning vevs have been already 
mentioned above. 
With  spontaneous breaking by 
a bulk scalar field one can introduce a  $\chi$ field 
that carries lepton number 2. 
It develops a small vacuum and gives mass to the neutrinos
which are generically of the form
\be 
m_\nu \sim \frac{\la H\ra^2}{M_\ast^2} \frac{\la\chi\ra_B}{M_\ast^{n/2}};
\ee
then, for n=2 and  
$M_\ast$ of the order of 100 TeV, we need $\la\chi\ra_B\sim (10~ GeV)^2$ to get 
$m_\nu\sim 10 ~eV$. 
Such small vacuums are possible in both this models, 
though it is  usually needed a small mass for $\chi$. 
Obviously, with Majorana masses, a bulk
neutrino is not needed but new physics must be invoked.
We should notice the there is also  a Majoron field associated to the
spontaneous breaking of the lepton number symmetry.
Its phenomenology depends on the details of the specific model.  
In the simplest scenario, the coupling $(LH)^2\chi$ 
is the one responsible for generating Majorana masses. 
It also gives  an important contribution for neutrinoless
double beta decay which is just right at the current  experimental 
limits~\cite{beta}.

\subsubsection{Models with local $B-L$ symmetry.}

Here we give an example  of a simple
model that uses the spontaneous breaking of a local $B-L$ symmetry
to generate neutrino masses~\cite{mp}. 
Consider a 5D model based in the gauge group
$SU(2)\times U(1)_I\times U(1)_{B-L}$, built, as usual, on the orbifold
$U(1)/Z_2$, with the matter content 
 \be 
 {\cal L}(2,0,-1)=\left(\ba{c} \nu\\ e\ea\right);~~ E(1,-1/2,-1)~.
 \ee
The scalar sector is chosen to contain  a doublet, $H(2,-1/2, 0)$, and a
singlet $\chi(1,1/2,1)$, which are used to break the symmetry down to 
the electromagnetic $U(1)_{em}$. Particularly $\la\chi\ra$ produce the
breaking of  $U(1)_{I}\times U(1)_{B-L}$ down to the hypercharge group
$U(1)_Y$. Also, electric charge is given by the linear
combination
 \be
 Q = T_3 + I + \frac{1}{2} (B-L) ~.
 \ee
As for the $Z_2$ parities of the fermion sector, we take the following fields
$ L={\cal L}_L,$ and $e_R = E_R,$ to be even, and thus   
${\cal L}_R$, and $E_L$ should be odd fields. Thus, at zero mode level we get 
usual SM lepton content. With this parities, since there is no right handed zero
mode neutrino, the theory has not  Dirac mass terms $\bar L H \nu_R$. 

Next we look for the simplest non-renormalizable operator that can generate
neutrino masses. It is the dimension 10 operator in 5D~\cite{mp,pp}
$({\cal L} H)^2 \chi^2$; 
which in the effective theory, after setting in the scalar vevs, 
generates the Majorana mass term
\be
\frac{h}{(M_\ast r)^2}~ \frac{(L H\chi)^2}{M_\ast^3}~.
\ee
If one takes $\la\chi\ra\sim 800~GeV$,  assuming that  
$M_\ast r\sim 100$ as suggested from the running of gauge couplings, 
and  $M_\ast \sim 100~TeV$ one easily gets a neutrino mass in the desired range,
$m_\nu\sim h\cdot~eV$.

 \subsection{Proton Stability in 6D models}

Keeping under control proton stability is more subtle. 
One might of course invoke global symmetries again, but this is less natural
since baryogenesis seems to require some degree of violation of the baryon
number, thus, without  the knowledge of the theory above $M_\ast$ 
it is difficult just to assume that  such operators are not being induced. 
  
In a recent paper~\cite{ponton} a  solution to the proton decay
problem was proposed in the context of the so called universal
extra dimension models (UED)~\cite{cheng2} where the number of
space-time dimensions where all standard model (SM) fields reside is six
and the fundamental scale of nature is in the TeV range. The main
observation of~\cite{ponton} is that in six
dimensional UED models, the extra space-time dimensions (the fifth
and sixth dimensions) provide a new $U(1)$ symmetry under which the
SM fermions are charged and enough of this symmetry survives
the process of orbifold compactification that it suppresses proton
decay to a very high degree. Besides, 6D SM have the remarkable property of
being and anomaly free theory only if the model contains a minimum of three
generations~\cite{poppitz}.

\subsubsection{Aspects of 6D SM.}

Let us  consider a six dimensional model~\cite{poppitz} based on the
standard gauge group $SU(3)_c\times SU(2)_L\times U(1)_{Y}$, with the  
following fermion content
\be 
{\cal{Q}}_{-}(3,2,1/3)~;~  
{\cal{L}}_{-}(1,2,-1)~;~
{\cal{U}}_{+}(3,1,4/3)~;~
{\cal{D}}_{+}(3,1,-2/3)~;~
{\cal{E}}_{+}(1,1,-2)~;~
N_{+}(1,1,0)~;
\label{6dfer}
\ee
where the numbers within parentheses are the gauge quantum numbers.
Here the subscripts $\pm$ denote the six dimensional chirality.
The corresponding six dimensional chirality projection operator is
defined as $P_\pm = \frac{1}{2}(1 \pm \Gamma^7)$, where $\Gamma^7$ is
itself given by the product of the six  Dirac
matrices already presented in
Eq.~(\ref{6dgammas}). As shown,  
these  are  are eight by eight matrices 
built out of  the well known $\gamma^\mu$ of the four dimensional
representation. In that representation one gets  
$\Gamma^7 = diag(1_{4\times 4},-1_{4\times 4})$

With the above assignments the model describes chiral interactions
that should be made anomaly free to be consistent.
There are two classes of  anomalies:
{\it local} and {\it global} anomalies
(for a discussion see~\cite{poppitz,borghini}).
Local anomalies are related to
infinitesimal  gauge and/or  coordinate transformations,
whereas global anomalies are essentially nonperturbative.

Each of above  fermion fields  is a four component field
with two 4-dimensional 2 component spinors with opposite 4D chirality
e.g. $Q_-$ has a left chiral $Q_{-,L}$ and a right chiral field
$Q_{-,R}$. As such the effective 4D theory is vector like at this
stage and we will need orbifold projections to obtain a chiral
theory. This is reflected in the fact that the theory contains no 
triangular anomalies.
In six dimensions, local anomalies arise from box one-loop diagrams
where the external legs are either gauge bosons or 
gravitons. Diagrams with only gauge bosons in the external legs
correspond to the pure gauge anomaly, whereas those with only gravitons
give the pure gravitational anomaly.
Diagrams with both gauge bosons and gravitons
correspond to mixed anomalies.

{\it Cancellation of local anomalies}.-
In the present  model $SU(3)$ is vector-like due to the replication of
representations with opposite chiralities. $U(1)$ and 
 $SU(3)_c\times U(1)_{B-L}$ anomalies do cancel
within each generation. Same holds for the subgroup $U(1)_Q$. In
fact, the model has no irreducible local gauge anomalies. The only
possible anomalies of this kind, which are $[U(1)]^4$ and
$[SU(3)]^3 U(1)$ vanish identically. 
All other anomalies associated
to: $[SU(2)]^4$; $[SU(3)]^2[SU(2)]^2$; and
$[SU(2)]^2 [U(1)]^2$; are reducible. They are not a matter of
concern, because they  can be canceled through the Green-Schwarz
mechanism~\cite{gs} by the introduction of an appropriate set of
two index antisymmetric tensors. The presence of reducible
anomalies is rather generic in six dimensional chiral theories,
thus, antisymmetric tensor are likely to be an ingredient of any
six dimensional model (see for instance the  models in
Refs.~\cite{cheng2,borghini,piai}). Notice that in fact all local
 gauge anomalies completely cancel. 

As the total number of fermions with chirality $+$ is equal to the
number of fermions with chirality $-$, there is no pure
gravitational anomaly. Regarding mixed anomalies, only those
associated to  diagrams with two gravitons in the external legs can
be non zero~\cite{gaume}. Again, such anomalies do vanish for 
$U(1)$ and $SU(3)$. 
Mixed anomalies that involve $SU(2)$ are all reducible, and
canceled by the same tensors that take care of the reducible pure
gauge anomalies.

{\it Global anomalies and the number of generations}.- 
Global anomalies are, on the other hand,
more restrictive for the fermion content of the model.
These anomalies are related to local symmetries that can not be deduced
continuously from the identity.
Cancellation of the of global gravitational anomalies  in six dimensions,
however, is
automatically insured by the cancellation of the local gravitational anomaly.
Therefore, only global gauge anomalies are  possible.
In general, they are associated to a non trivial topology of the gauge group.
Particularly, they arise in six dimensional theories
when the sixth homotopy group, $\pi_6$,
of the gauge groups  is non trivial.
Cancellation of such an anomaly needs an appropriate matter content.
As a matter of fact, they may occur for $SU(3)$
as well as $SU(2)$ gauge theories~\cite{witten2,vafa}.
Given that $\pi_6[SU(3)]= Z_{6}$  and $\pi_6[SU(2)]= Z_{12}$,
the  cancellation of the  global gauge anomalies constrains the number of
chiral triplet color  representations
in the model, $N_c(3_\pm)$, to satisfy:
 \be
 N_c(3_+) - N_c(3_-) = 0 ~~\mbox{mod}~6 ~
 \ee
As $SU(3)$ is vector like  this condition is naturally fulfilled.
For the number of $SU(2)$ chiral doublets, $N(2_\pm)$,
it also  requires that
 \be
 \label{cons}
 N(2_+) - N(2_-) = 0 ~~\mbox{mod}~6 ~.
 \ee
Last condition indicates that the global anomaly does not cancel
within a single family, because all doublets  are all of the same chirality.
One easily sees that the above constraint can
be written in a unique way in terms of the number of generations,
$n_g$, which is the number of exact replications of our matter
content,  as follows~\cite{poppitz}
 \be
 n_g =0~~\mbox{mod}~ 3~.
  \ee
Hence, 3 is the minimum number of generations for which the theory
is mathematically consistent. 
This is a remarkable result that survives even in some extensions, as in
some left-right models~\cite{6dlr}, which also account for the generation of 
neutrino mass with the same fermion content.

\subsubsection{Lorentz invariance and Baryon non conservation.}

As in the standard 4D theory, to consider which processes are  possible
we will require all renormalizable and non
renormalizable operators to obey all symmetries of the theory. 
Apart from SM gauge transformations, 
the theory now has to be invariant under a larger
Lorentz group given by  $SO(1,5)$. Of course usual 4D Lorentz
transformations are  included in this extended Lorentz
group as the  $SO(1,3)$ subgroup. If we 
denote the six space-time coordinates by $(x^0,x^1, x^2, x^3, x^4, x^5)$, 
we can identify the transformation group associated to pure rotations 
in the plane formed by the fifth and sixth extra dimensions
by  $U(1)_{45}$, which is
contained in $SO(1,5)$. The generator of this group in the representation given
above [see Eq.~(\ref{6dgammas})] is 
$\Sigma_{45}=\frac{i}{2}[\Gamma^4 , \Gamma^5] \equiv
\gamma_5 \otimes \sigma_3$. Since $\Sigma_{45}$ is diagonal it is a 
well defined
quantum number for the left and right components of the fermions. In general, 
one can easily see that 
\be
\Sigma_{45}\Psi_\pm = \pm \gamma_5 \Psi_\pm~.
\ee
Hence  4D chiral parts have indeed an explicit $\Sigma_{45}$
charge. Particularly  we get that  $\Psi_{+R}$ and $\Psi_{-L}$ have 
both $\Sigma_{45}=1$. Looking at the fermion sector given in Eq.~(\ref{6dfer}),   
we straightforwardly identify this components as those containing 
the zero mode part of the theory
such that the matter content  at low energy would be precisely that of the SM.
Thus,  $\Sigma_{45}=-1$ fields should only appear at the excited level.

That all SM fields are equally charged under  $\Sigma_{45}$ has 
important consequences for the non renormalizable operators 
responsible of proton decay. Consider for instance  
$\bar Q^c Q\bar Q^c L$, where 6D charge conjugation is defined by 
$\Psi^c=C\bar\Psi^T$, where $C=\Gamma^0 \Gamma^2 \Gamma^4$.
One immediately notice that 
this operator has $\Delta \Sigma_{45}=4$, thus it is 
non Lorentz invariant in 6D, and so it is forbidden. 
Clearly, all operators of this sort have same fate.
The first non renormalizable operators that may account for baryon number
violation one can write are dimension 16 operators~\cite{ponton}, as
$\left(\bar{\cal L}{\cal D}\right)(\bar{\cal Q}^c N)(\bar{\cal D}^c \nt{D} N)$.
Notice the operator involve only bilinears that are  Lorentz
invariant already, and since for proton decay one needs to involve at least
three quarks, six different fermions are needed to built up the operator.
Therefore the simplest proton decay processes should involve three leptons in
the final states (see references~\cite{ponton,6dlr} for some explicit examples).
For instance, 
for the new nucleon decay mode 
$N\rightarrow \pi \nu_e \nu_s  \nu_s$ one estimates the
life time to be
 \be 
 \tau_p \approx 6\times 10^{30}~{\rm yr}
 \cdot\left[{10^{-4}\over\Phi_n}\right]
 \left({\pi rM_*\over 10}\right)^{10}~
 \left({M_*\over 10~{\rm TeV}}\right)^{12}~,
 \ee
which is large enough as to be consistent with the experiment 
even with a fundamental scale as low as 10 TeV~\cite{pdg}. 
Above,  $\nu_s$ stands for the sterile neutrino contained in $N_+$;
and  we have explicitly introduced the  contribution of the kinematical
phase space factor, $\Phi_n$, which depends in the specific process with
$n$ final states. Also a possible order one form factor which enters in
the case of two pion production has not been written. As usual, $r$ represents
the size of the compact space. 
Among other baryon number violating processes, 
there is also an intriguing invisible decay for the neutron 
$n\rightarrow \nu_e \nu_s \nu_s$~\cite{ponton,6dlr}, 
that is still above current experimental limits
for which a life time $\tau_n>2\cdot 10^{29}$~yrs. applies~\cite{snon}.
Other interesting modes would be
$p\rightarrow e^-\pi^+\pi^+\nu_s\nu_s$; and $n\rightarrow e^-\pi^+\nu_s\nu_s$.
For comparison, searches for the decays
$p\rightarrow e^-\pi^+\pi^-$; and $n\rightarrow e^-\pi^+$ set
limits in about $\tau_p>3\cdot10^{31}$~yrs.~\cite{frejus} and
$\tau_n>6.5\cdot10^{31}$~yrs.~\cite{seidel} respectively. 

\begin{figure}[ht]
\includegraphics[scale=.8]{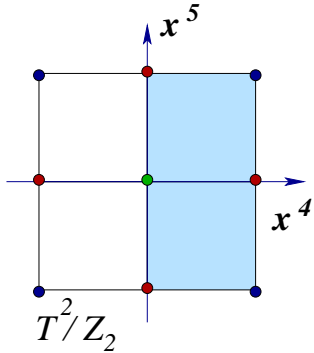}\hspace{2pc}%
\begin{minipage}[b]{18pc}
\caption{\label{t2z2}Fixed points of the $Z_2$ orbifolding of the torus, here
represented by the whole squared (of size $r$) 
in the $x^4$-$x^5$ plane. 
The shadowed region corresponds to the actual fundamental space.}
\end{minipage}
\end{figure}

It is certainly impressive that the extra symmetries of the space are just
enough as to provide us with an understanding of proton stability. 
However, the argument has a weakness. It relies on a symmetry
which is usually  broken by compactification. It is then important to 
know up to what extent the argument holds on compact space. 
It actually does, provided the rotational symmetry $U(1)_{45}$ 
is not completely broken.
Consider for instance the orbifold $T^2/Z_2$, where $T^2$ is the torus and 
$Z_2$ the identification $\vec{y}\rightarrow -\vec{y}$, where 
$\vec{y} =(x^4,x^5)$. As it can be seen from 
Fig.~\ref{t2z2}, where we have represented the physical compact 
space on the covering space ${\cal R}^2$, 
the compactification
breaks the $SO(1,5)$ group down to $SO(1,3)\times {\cal{Z}}_4$, where 
${\cal Z}_4$ is the group of discrete rotations in the $x^4-x^5$ plane around
the origin by $\pi/2$, a subgroup of $U(1)_{45}$. 
Clearly, this rotation maps fixed points into themselves.
For fermions ${\cal{Z}}_4$ rotations  become ${\cal{Z}}_8$ rotations. 
Any  operator in the  effective theory should be invariant 
under these transformations generated by the same $\Sigma_{45}$ matrix. 
Therefore, any  fermionic operator should satisfy the selection rule
 \be 
 \Delta\Sigma_{45}=0~~ mod~ 8~.
 \ee
Usual dimension 10 operators $LQQQ$ (dimension 6 in 4D), do not fulfill this
rule, and thus, they remain forbidden. The dimension 16 operators mentioned
above do remain, and so does a suppressed proton decay.

It is worth noticing that  because charge conjugation operator, $C$, 
is such that it commutes with 6D chiral operator, $\Gamma_7$, which implies that 
$(\Psi_\pm)^C=(\Psi^C)_\pm $, and also because $C$
anticommutes  with $\Sigma_{45}$, 
there are not Majorana mass terms for neutrinos in these theories.
The neutrino should rather be a Dirac field. 
In the SM presented above, if $N_{R}$ is the even part of the field, then 
Dirac mass terms, $\bar{L} H N_R$, are indeed possible. Smallness is, however,
unnatural. One should relay in very small Yukawa couplings. On the other hand,
if one rather choose $N_{R}$ to be and odd field, so that $N_{L}$ is even, 
and extend the gauge sector to contain $B-L$, a solution may be at
hand~\cite{6dlr}. In such a case $\bar{L} H N_R$  does not give neutrino
masses ($N_R$ has only KK modes), but these may be introduced via non
renormalizable operators.

\subsection{Split Fermions. Hierarchies without Symmetries}

Another  interesting 
mechanism that explain how proton decay could get  suppressed at the
proper level appeared in~\cite{sch1}.  It relays on the idea that the
branes are being formed from an effective mechanism that traps the SM
particles in it, resulting in a  wall with thickness $L\sim M_\ast^{-1}$, 
where the fermions are stuck at different points.  Then, fermion-fermion
couplings get suppressed due to the exponentially small  overlaps of their
wave functions. This provides a framework for understanding both the
fermion mass hierarchy and proton stability without imposing extra 
symmetries, but rather in terms of a higher dimensional
geography~\cite{sch2}.   Note that the dimension where the gauge fields
propagate does not need to be orthogonal to the millimetric
dimensions, but  gauge  fields may be restricted to live in a smaller
part of that extra  dimensions.  Here we briefly summarize those ideas.

\subsubsection{Localizing wave functions on the brane.}

Let us start by assuming that  translational invariance along the fifth
dimension is being broken
by  a bulk scalar field $\Phi$
which develops a spatially varying expectation value $\la\Phi\ra(y)$. We
assume that this expectation
value have the shape of a domain wall transverse to the extra
dimension and is centered at $y=0$. With this background 
a bulk fermion will have a zero mode that is stuck at the zero of 
$\la\Phi(y)\ra$. To see this let us consider the action
 \be 
 S= \int d^4x\ {\rm d}y\,
 \overline\Psi\left[i\, \Gamma^M\!\partial_M +
 \la\Phi\ra(y)\right] \Psi ,
\label{single5}
 \ee 
in the chiral basis given by Eqs.~(\ref{5dgammas}) and~(\ref{5dfer}) . 
By introducing the expansions
 \be
  \psi_L(x,y) = \sum_n f_n(y)\psi_{nL}(x);\qquad \mbox{ and } \qquad
 \psi_R(x,y) = \sum_n g_n(y)\psi_{nR}(x);
 \ee
where $\psi_{n}$ are four dimensional spinors, we get for the 
$y$-dependent functions $f_n$ and $g_n$ the equations
 \be
 \left( \partial_5 +\la\Phi\ra\right) f_n  + m_n g_n= 0;
 \qquad \mbox{ and } \qquad
 \left(-\partial_5+\la\Phi\ra \right) g_n + m_n f_n= 0;
 \ee
where now $m_n^2 = p_\mu p^\mu$ stands for the 4D mass parameter. 
Therefore, the zero modes have the profiles~\cite{sch1}
 \be
 f_0(y)\sim \exp\left[-\int^{y}_0\! ds \la\Phi\ra(s) \right]
 \qquad \mbox{ and }\qquad
 g_0(y) \sim \exp\left[\int^{y}_0 ds \la\Phi\ra(s)\right];
 \ee
up to normalization factors. Notice that when the extra space is supposed
to be finite, both modes are normalizable. For the special choice
$\la\Phi\ra(y)= 2\mu_0^2y$, we get $f_0$ centered at $y=0$ with the gaussian
form
 \be 
 f_0(y)= \frac{\mu_0^{1/2}}{(\pi/2)^{1/4}}\ \exp\left[{-\mu_0^2{y}^2}\right] .
 \ee
The other mode has been
projected out from our brane by being pushed away to the end of the space.
Thus, our theory in the wall is a chiral theory. 
Notice that a negative coupling among $\Psi$ and $\phi$ will instead
project out the left handed part.

The generalization of this technique to the case of several fermions is
straightforward. The action (\ref{single5}) is generalized to
 \be 
 S = \int\! d^5x\, 
 \sum_{i,j}\bar\Psi_i[i\,\Gamma^M\partial_M + \lambda\la\Phi\ra-m]_{ij} \Psi_j\ ,
 \label{multi5}
 \ee
where general Yukawa couplings $\lambda$ 
and other possible five dimensional masses $m_{ij}$
have been considered. For simplicity we will assume both terms diagonal.
The effect of these new parameters is a shifting of the wave functions,
which now are centered around the zeros of $\lambda_i\la\Phi\ra-m_i$.
Taking $\lambda_i =1$ with the same profile for the vacuum leads to
gaussian distributions centered at $y_i = m_i/2\mu_0^2$. Thus, at low
energies, the above action will describe a set of non interacting four
dimensional chiral fermions localized at different positions in the fifth
dimension.

Localization of gauge and Higgs bosons  needs extra assumptions. The
explanation of this phenomena is close related with the actual way the
brane was formed. A field-theoretic mechanism for localizing gauge fields
was proposed by Dvali and Shifman~\cite{shifman}
and was later extended and applied  in
\cite{dvali}. There, the idea is to arrange for the gauge group to confine
outside the wall; the flux lines of any electric sources turned on inside
the wall will then be repelled by the confining regions outside and forced them 
to propagate only inside the wall. This traps a massless gauge field on
the wall. Since the gauge field is prevented to enter the confined region,
the thickness $L$ of the wall acts effectively as the size of the extra 
dimension in which the gauge fields can propagate. In this
picture, the gauge couplings will exhibit power law running
above the scale $L^{-1}$.

\subsubsection{Fermion mass hierarchies and proton decay.}

Let us consider the Yukawa coupling among the Higgs field and the leptons:
$\kappa H  L^{T}E^c$; where the massless zero
mode  $l$ from $L$ 
is localized at $y=0$ while $e$ from 
$E^c$ is localized at $y=r$. Let us also assume that the Higgs zero mode is
delocalized inside the wall. Then the zero modes term of this coupling will
generate the effective Yukawa action
 \be 
  S_{Yuk} = \int\! d^4 x\, \kappa\, h(x) l(x) e^c(x)\ 
  \int\! dy\ \phi_l(y)\ \phi_{e^c}(y)\ ,
 \ee
where $\phi_l$ and $\phi_{e^c}$ represent the gaussian profile of the
fermionic modes. Last integral gives the overlap of the wave functions,
which is exponentially suppressed~\cite{sch1} as
 \be 
  \int\! dy\ \phi_l(y)\ \phi_{e^c}(y)\ = e^{-\mu_0^2r^2/2}.
 \ee
This is a generic feature of this models. The effective coupling of any
two fermion fields is exponentially suppressed in terms of their
separation in the extra space. Thus, the  explanation for the mass
hierarchies becomes a problem of the cartography on the extra
dimension. A more detailed analysis was presented in~\cite{sch2}.

Let us now show how  a fast proton decay 
is evaded in these models.  
Assume, for instance,   that all quark fields are localized at 
$y=0$ whereas the leptons are at $y=r$. 
Then, let us consider the following baryon
and lepton number violating operator
\be
S \sim \int d^5x\,  \frac{(Q^{T} C_5 L)^{\dagger} (U^{c T}
C_5 D^c)}{M^3} .
\ee
In the four dimensional effective theory, once we have 
introduced the zero mode
wave functions, we get the suppressed action~\cite{sch1}
\be
S \sim \int d^4x\,\, \lambda \times \frac{(q l)^{\dagger}
(u^c d^c)}{M^2}
\ee
where $\lambda \sim \int dy\, \left[ e^{-\mu_0^2y^2}\right]^3
e^{-\mu_0^2(y-r)^2}\sim e^{-3/4 \mu_0^2r^2}$.
Then,  for a separation of $\mu_0 r =10$ we obtain $\lambda\sim10^{-33}$
which renders these operators completely safe even for $M_\ast\sim 1$ TeV.
Therefore, we may 
imagine a picture where quarks and leptons are localized near
opposite ends of the wall so that $r \sim L$.
This mechanism, however, does not work for suppressing the another
dangerous operator
$(LH)^2/M$ responsible of a  large neutrino mass.

\section{Warped Extra Dimensions}

So far we have been working in the simplest picture where the energy density on
the brane does not affect the space time curvature, but rather it has
been taken as a perturbation on the flat extra space. 
However, for large brane densities this
may not be the case. 
The first approximation to the problem can be done by 
considering a five dimensional model where branes
are located at the two ends of a closed fifth dimension. 
Clearly, with a single extra dimension, the gravity flux produced by a single
brane at $y=0$ can not softly close into itself 
at the other end the space, making the model unstable, just as a charged
particle living in a closed one-dimensional world does not define a stable
configuration. Stability can only be
insured by the introduction of a second charge (brane).
Furthermore, to balance branes energy and still get 
flat (stable) brane metrics, one has to compensate the effect on the space by
the introduction of a
negative cosmological constant on the bulk. Hence, the fifth dimension would be 
a slice of an Anti de-Siter space with flat branes at its edges. 
Thus, one can keep the branes flat paying the
price of curving the extra dimension. Such curved extra dimensions are usually
referred as warped extra dimensions. 
Historically, the possibility was first 
mentioned by Rubakov and Shaposhnikov in Ref.~\cite{rubakov}, 
who suggested that the cosmological constant problem could be understood under
this light: the matter fields vacuum energy on the brane could be canceled
by the bulk vacuum, leaving a zero (or almost zero) cosmological constant for
the brane observer. No specific model was given there, though. 
It was actually  Gogberashvili~\cite{merab} who
provided the first exact solution for a warped metric, nevertheless, 
this models are best known after Randall and Sundrum (RS) 
who linked the solution to the 
the hierarchy problem~\cite{rs1}. Later developments suggested that the warped
metrics could even provide an alternative to compactification for the extra
dimensions~\cite{rs2,nimars}. In what follows we shall discuss a concrete
example as presented by Randall and Sundrum.

\subsection{Randall-Sundrum Background and the Hierarchy Problem}
\subsubsection{Randall-Sundrum background.}

Lets consider the following setup. A five dimensional space with an
orbifolded fifth dimension of radius $r$ and coordinate $y$ which takes values
in the interval $[0,\pi r]$.
Consider two branes at the fixed (end) points
$y=0,\pi r$; with tensions $\tau$ and $-\tau$ respectively. 
For reasons that should become clear later on,  the brane at $y=0$
($y=\pi r$) is usually called the hidden (visible) or Planck (SM) brane. 
We will also assign to  the bulk  a negative cosmological constant $-\Lambda$. 
Contrary to our previous philosophy, here we shall 
assume that all parameters are of the
order of the Planck scale. 
Next, we ask for the solution that gives a 
flat induced metric on the branes such that 4D Lorentz invariance is respected.
To get a consistent answer, one has to require that 
at every point along the fifth dimension the induced metric should be the
ordinary flat 4D Minkowski metric. Therefore, the components of the 5D metric
only depend on the fifth coordinate. Hence, one gets the ansatz
 \be 
  ds^2 = g_{AB} dx^A dx^B = 
  \omega^2(y)\eta_{\mu\nu}dx^\mu dx^\nu - dy^2~,
 \label{wmetric}
 \ee
where we parameterize $\omega(y) = e^{-\beta(y)}$.
The metric, of course, can always be written in different coordinate systems.
Particularly, notice that one can easily go to the conformally 
flat metric, where
there is an overall factor in front of all coordinates,  
$ds^2 =  \omega^2(z)[\eta_{\mu\nu}dx^\mu dx^\nu - dz^2]$, where the new
coordinate $z$ is a function of the old coordinate $y$ only.

Classical action  contains $S= S_{grav} + S_{h} + S_{v}$; where
 \be
 S_{grav} = \int \! d^4x\, dy \sqrt{g_{(5)}} \left( \frac{1}{2k_\ast^2} R_{5}
 +\Lambda\right) ~,
 \ee
gives the bulk contribution, whereas the visible and hidden brane 
parts are given by
 \be 
 S_{v,h}= \pm~\tau~\int\! d^4x \sqrt{-g_{v,h}}~,
 \ee
where $g_{v,h}$ stands for the induced metric at the visible and hidden
branes, respectively. Here $\kappa_\ast^2 = 8\pi G_\ast = M_\ast^{-3}$.

Five dimensional  Einstein equations for the given action become
 \be
 G_{MN} =  -k_\ast^2\Lambda\,g_{MN} + 
   k_\ast^2\tau\, \sqrt{\frac{-g_{h}}{g_{(5)}}} 
    \delta_M^\mu \delta_N^\nu g_{\mu\nu}\delta(y) 
      \nonumber  -  k_\ast^2\tau\, \sqrt{\frac{-g_{v}}{g_{(5)}}} 
    \delta_M^\mu \delta_N^\nu g_{\mu\nu}\delta(y-\pi r) 
\label{eers}
 \ee
where the Einstein tensor $G_{MN} = R_{MN} - \frac{1}{2} g_{MN} R_{(5)}$ as
usual.  They are easily reduced into two simple independent equations. 
First, we can expand
the $G_{MN}$ tensor components  using the metric
ansatz (\ref{wmetric}) to show that
 \be
 G_{\mu\nu} = -3\, g_{\mu\nu}\left( -\beta'' + 2 (\beta')^2 \right)~; \quad
 G_{\mu 5} =0~; \quad \mbox{ and } \quad
 G_{55} = -6\, g_{55} (\beta')^2  ~.
 \ee
Next, using the RHS of Eq.~(\ref{eers}), one gets that 
 $6 (\beta')^2 =  k_\ast^2  \Lambda$, 
and 
\be 
 3\beta'' = k_\ast^2  \tau\, \left[\delta(y) -\delta(y-\pi r)\right]. 
\ee
Last  equation, clearly,  defines the boundary conditions for the function
$\beta'(y)$ at the two branes (Israel conditions). 
Clearly, the solution is  $\beta(y) = \mu |y|$, where 
 \be 
  \mu^2 =\frac{k_\ast^2\Lambda}{6}= \frac{\Lambda}{6 M_\ast^3}~,
 \label{rsmu}
 \ee
with the subsidiary fine tuning condition 
 \be 
 \Lambda = \frac{\tau^2}{6 M_\ast^3}~,
 \label{rsfine}
 \ee
obtained from the boundary conditions,  
that is equivalent to the exact cancellation of the effective four
dimensional cosmological constant. The background metric is therefore
 \be 
  ds^2 = e^{-2\mu |y|}\eta_{\mu\nu}dx^\mu dx^\nu - dy^2~.
 \label{rsmetric}
 \ee
The effective Planck scale in the theory is then given by 
 \be
 M_{P}^2 = \frac{M_\ast^3}{\mu}\left(1- e^{-2\mu r\pi}\right). 
 \label{rsmp}
 \ee
Notice that for large 
$r$, the exponential piece becomes negligible,
and above expression has the familiar form given in Eq.~(\ref{mp}) for one
extra dimension of (effective) size $1/\mu$.

\subsubsection{Visible versus Hidden Scales Hierarchy.} 

The RS metric has a peculiar feature. Consider a given distance, $ds_0^2$,
defined by fixed intervals $dx_\mu dx^\mu$ from brane coordinates.
If one maps the interval from hidden to visible
brane, it would appear here 
exponentially smaller than what is measured at the
hidden brane, i.e., $ds_0^2|_{v} = \omega^2(\pi r) ds_0^2|_{h}$. 
This scaling property 
would have interesting consequences when introducing fields to live on any
of the branes. Particularly, let us discuss what happens for a theory 
defined on the visible brane.

The effect of RS background on visible brane field parameters 
is non trivial. Consider for instance the scalar field  
action for the visible brane at the end of the space  given by
 \[
 S_H =
\int\! d^4x\, \omega^4(\pi r) \left [\omega^{-2}(\pi r)
     \partial^\mu H\partial_\mu H - \lambda\left(H^2 - 
     \hat v_0^2\right)^2\right].
 \]  
As a rule, we choose all dimensionful parameters 
on the theory to be naturally given in terms of  $M_\ast$, and this 
to be close to $M_P$. 
So we take  $\hat v_0\sim M_\ast$.
After introducing the normalization 
$H\rightarrow \omega^{-1}(\pi r) H = e^{\mu r\pi} H$ to
recover  the canonical kinetic term, the above action becomes
 \be
 S_H = \int\! d^4x   
 \left[ \partial^\mu H\partial_\mu H - \lambda\left(H^2 - 
     v^2\right)^2\right],
 \ee
where the actual vacuum $v= e^{-\mu r\pi}\hat v_0$. Therefore, by choosing 
$\mu r \sim 12$, the physical mass of the scalar field, and its vacuum,
would naturally appear at the TeV scale rather than at the Planck scale, 
without the need of any large hierarchy on the radius~\cite{rs1}. 
Notice that, on the contrary, any field
located on the other brane will get a mass of the order of $M_\ast$. 
Moreover,  it also implies that no  particles exist in the visible brane with
masses larger than TeV. This observation has been consider a nice 
possible way of solving the scales hierarchy problem. For this reason, the
original model proposed that our observable Universe resides on the brane
located at the end of the space, the visible brane. So the other brane 
really becomes hidden. This two brane model is sometimes called RSI model.

 \subsection{KK Decomposition on RS}
 
As a further note, notice that since there is everywhere 4D Poincar\'e
invariance, every bulk field on the RS background can be expanded into four
dimensional plane waves $\phi(x,y)\propto e^{ip_\mu x^\mu} \varphi(y)$. This
would be  the basis for the Kaluza Klein decomposition, that we shall now
discuss.
Note also that the
physical four momentum of the particle at any position of the brane goes as 
$p_{phys}^\mu (y) = \omega^{-1}(y) p^\mu$. Therefore, modes which are soft on
the hidden brane, become harder at any other point of the bulk. 

Lets consider again a bulk scalar field, now on the RS background metric. The
action is then
 \be 
 S[\phi] = \frac{1}{2}\int\!d^4x\, dy \sqrt{g_{(5)}}
 \left(g^{MN}\partial_M\phi \partial_N\phi - m^2\phi^2 \right)~.
 \ee
By introducing the factorization $\phi(x,y) = e^{ip_\mu x^\mu} \varphi(y)$ 
into the equation of motion, one gets that the KK modes satisfy
 \be 
 \left[ - \partial_y^2 + 4\mu~ {\rm sgn}(y) \partial_y + m^2 +
 \omega^{-2}(y)\, p^2 \right] \varphi(y) =0~,
 \label{kkrs}
 \ee
where $p^2= p^\mu p_\mu$ can also be interpreted as the effective four
dimensional invariant mass, $m_n^2$. It is possible, through a functional  
re-parameterization and a change of variable, 
to show that
the solution for $\varphi$ can be written in terms of Bessel functions
of index $\nu = \sqrt{4 + m^2/\mu^2}$~\cite{gw0,Rubakov2}, as
follows 
 \be 
 \varphi_n(y) = \frac{1}{ N_n \omega^2(y)}\, 
 \left[ J_\nu \left( \frac{m_n}{\mu\omega(y)}\right) 
 + b_{n\nu} Y_\nu \left( \frac{m_n}{\mu\omega(y)}\right) \right]\,,
 \label{kkchi}
 \ee 
where $N_n$ is a normalization factor, $n$ labels the KK index, and the 
constant coefficient $b_{n\nu}$ has to be fixed by 
the continuity conditions at one of the boundaries. The other boundary
condition would serve to quantize the spectrum. For more details the reader
can see
Ref.~\cite{gw0}. Here we will just make some few comments about. 
First, for $\omega(\pi r)\ll 1$, 
the discretization condition that one gets for 
$x_{n\mu} = m_n/\mu\omega(y)$  looks as
 \be 
 2 J_\nu(x_{n\mu}) + x_{n\nu} J'_\nu(x_{n\mu}) =0~.
 \label{kkmrs}
 \ee
Therefore, the lowest mode satisfies $x_{1\mu}\sim {\cal O}(1)$, which means
that  $m_1 \simeq \mu e^{-\mu r \pi}$. For the same range of parameters we
considered before to solve the hierarchy problem, one gets that lightest KK
mode would have a mass of order TeV or so.
Next, for the special case of a originally massless field ($m=0$),
one has $\nu=2$, and thus the first solution to Eq.~(\ref{kkmrs}) is just
$x_{12}=0$, which indicates the existence of a massless mode in the spectrum.  
The next zero of the equation would be of order one again, thus the KK tower
would start at $\mu e^{-\mu r \pi}$. The spacing among two consecutive KK
levels would also be of about  same order. There is no need to stress that
this would actually be the case of the graviton spectrum.
This makes the whole spectrum
completely distinct from the former ADD model. With such heavy graviton modes 
one would not expect to have visible deviations on the short distance gravity
experiments, nor constrains from BBN, star cooling or astrophysics in general. 
Even colliders may not be able to test direct production or exchange of such
heavy gravitons.

\subsection{Radius Stabilization}

The way RSI model solves the hierarchy problem between $m_{EW}$ and $M_P$ 
depends on the interbrane spacing $\pi r$. Stabilizing the bulk
becomes in this case an important issue if one is willing to keep this
solution. The dynamics of the extra dimension would give rise to a 
running away radion field, similarly  as it does for the ADD case.
A simple exploration of the metric (\ref{rsmetric}), 
by naively setting a slowly time dependent bulk radius $r(t)$, shows that 
 \be 
 ds^2 \rightarrow e^{-2\mu r(t) |\phi|} \eta_{\mu\nu} dx^\mu dx^\nu - 
 r^2(t) d\theta^2~;
 \ee
with $\theta$ the angular coordinate on the half circle $[0,\pi]$. This suggest  
that if the interbrane distance changes the visible brane 
expands (or contracts) exponentially. 
The radion field associated to the fluctuations of the radius, 
$b(t) = r(t) - r$, is again massless
and thus it violates the equivalence principle.  
Moreover, without a stabilization mechanism for the radius, our brane
could expand forever. Some early discussions on this and other 
issues can be found in Refs.~\cite{gw,rsradion,rrs1}. 

A simple and  elegant solution for stabilization in RSI 
was proposed by 
Goldberger and Wise~\cite{gw}. The central  idea is as follows: 
if there is a vacuum
energy on the bulk, whose configuration breaks translational invariance along
fifth dimension, say  $\langle E \rangle (y)$, then, the effective four
dimensional theory would contain a radius dependent potential energy
 \[ V(r) = \int\! dy\, \omega^4(y)\, \langle E \rangle (y) ~.\]
Clearly, if such a potential has a non trivial minimum, stabilization would be
insured. The radion would feel a force that tends to keep it at the
minimum.
The  vacuum energy $\langle E \rangle (y)$ may come from many sources.
The simplest possibility one could think up on is a vacuum induced by a bulk
scalar field, with non trivial boundary conditions, 
 \be 
 \langle \phi \rangle (0) = v_h \qquad \mbox{ and } \qquad 
 \langle \phi \rangle (\pi r) = v_v~.
 \ee
The boundary conditions would amount for a non trivial profile of 
$\langle \phi \rangle (y)$ along the bulk. Such boundary conditions may arise,
for instance, if $\phi$ has localized interaction terms on the branes, as
$\lambda_{h,v} (\phi^2 - v_{h,v}^2)^2$, which by themselves develop non zero
vacuum expectation values for $\phi$ located on the branes. 
The vacuum is then the $x$ independent solution 
to the equation of motion (\ref{kkrs}), which can be written as 
$ \langle \phi \rangle (y) = \omega^{-1}(y)
  \left[ A \omega^{-\nu}(y) + B \omega^{\nu}(y)\right]$,
where $A$ and $B$ are constants to be fixed by the boundary conditions.
One then obtains the effective 4D vacuum energy 
 \be 
 V_\phi(r) =  \mu (\nu+2) A^2 \left(\omega^{-2\nu}(\pi r) - 1\right) 
   + \mu (\nu-2) B^2 \left(1 - \omega^{2\nu}(\pi r) \right)
 \ee
After a lengthly calculation, in the limit where $m\ll \mu$
one finds that above potential has a non trivial minimum for 
 \be 
 \mu r = \left(\frac{4}{\pi}\right) \frac{\mu^2}{m^2} \ln
 \left[\frac{v_h}{v_v}\right]~.
 \ee
Hence, for $\ln(v_h/v_v)$ of order one, the stable value for the radius 
goes proportional to the curvature parameter, $\mu$, and inversely to the squared
mass of the scalar field. Thus, one only needs that $m^2/\mu^2\sim 10$ to get 
$\mu r\sim 10$, as needed for the RSI model.

One can get a bit suspicious about whether the vacuum energy 
$\langle \phi \rangle (y)$ may disturb the background metric. 
It actually does, although the correction is negligible as the 
calculations  for the Einstein-scalar field coupled equations 
may show~\cite{gw,rrs1}.

\subsection{RSII: A Non Compact Extra Dimension}

The background metric solution (\ref{rsmetric}) does not actually need 
the presence of the negative tension brane to hold as an exact solution 
to Einstein equations. Indeed the warp factor $\omega(y) = e^{-\mu |y|}$ 
has been  determined only by the Israel conditions at the $y=0$ boundary. That
is, by using   $\omega'' = \mu^2 \omega - \mu \omega \delta(y)$ in Einstein
equations, 
which implies equations (\ref{rsmu}) and (\ref{rsfine}). 
It is then tempting to `move' the 
negative tension brane to infinity, which  renders a non compact  fifth
dimension. The picture becomes esthetically more appealing, 
it has no need for compactification. 
Nevertheless, one has to ask now the question of whether such
a possibility is at all consistent with observations.  
It is clear that the Newton's constant is now simply
 \be 
 G_N = \mu G_\ast~
 \label{rsgn}
 \ee
--just take the limit $r\rightarrow \infty$ in Eq.~(\ref{rsmp})--, 
this reflects
the fact that although the extra dimension is infinite, 
gravity  remains  four dimensional at large distances (for $\mu r\gg 1$). 
This is, 
in other words, only a consequence of the flatness of the brane. 
We shall expand our discussion on this point in following sections.
Obviously, with this setup, usually called the RSII model~\cite{rs2}, 
we are giving up the
possibility of explaining  the hierarchy between Planck and electroweak scales.
The interest on this model remains, however, due to potentially 
interesting physics at  low energy, and also due to  its connection to the
AdS/CFT correspondence~\cite{ads/cft}.

Although the fifth dimension is infinite, the point $y=\infty$ is in fact a
particle horizon. Indeed, the first indication comes from the metric, since 
$\omega(y\rightarrow\infty) = 0$. The confirmation would come from considering 
a particle moving away from the brane on the geodesics
$y_g(t) = \frac{1}{2\mu} \ln(1+\mu^2 t^2)$~\cite{ruba}. 
The particle accelerates towards
infinity, and its velocity tends to speed of light. The proper time interval is
then 
 \be 
 d\tau^2 = \omega^2(y_g(t)) dt^2 - \left(\frac{dy_g}{dt}\right)^2 dt^2~.
 \ee
Thus, the particle reaches infinity at infinite time $t$, but 
in a finite proper time $\tau = \pi/2\mu$.

\subsection{Graviton Localization}

In order to understand why gravity on the brane remains four dimensional at
large distances, even though the fifth dimension is non compact, one has to
consider again the KK decomposition for the graviton modes, with particular
interest on the shape for  the zero mode wave function~\cite{rs2}. 
Consider first the
generic form of the perturbed background metric
 \[ ds^2 = \omega^2(y)g_{\mu\nu}dx^\mu dx^\nu + A_\mu dx^\mu dy - b^2 dy^2~.\]
Due to the orbifold projection $y\rightarrow -y$, the vector component $A_\mu$
has to be odd, and thus it does not contain a zero mode. Therefore at the zero
mode level only the true four dimensional graviton  and the scalar (radion)
should survive. Let us concentrate on the 4D graviton perturbations only.
Introducing the small field expansion as 
$g_{\mu\nu} = \eta_{\mu\nu} + \omega^{-2}h_{\mu\nu}$, and using the 
gauge fixing conditions $\partial_\mu h^\mu_\nu =0 = h^\mu_\mu$, 
one obtains  the wave equation 
 \be 
 \left[\partial_y^2  - 4\mu^2  - 
 \frac{m^2}{\omega^2(y)}  - 4\mu\delta(y)\right]h=0~;
 \ee
where the Lorentz indices should be understood. In the above equation the mass
$m^2$ stands for the effective four dimensional mass $p^\mu p_\mu = m^2$.
It should be noticed that the mass spectrum would now be continuous and 
starts at $m=0$. In this situation the KK are normalized to a delta function,
$\int\! dy~ \omega^{-2}(y)\, h_m(y)\, h_{m'} = \delta(m-m')$.
 
Introducing the functional re-parameterization
$z = sgn(y)\, \left(\omega^{-1}(y) -1\right)/\mu$
and 
$\Psi(z) = \omega^{-1/2}(y)~ h(y)$;
one can write the equation of motion for the KK modes as the Scrh\"odinger
equation~\cite{rs2}
\be 
\left[ -\frac{1}{2} \partial_z^2 + V(z)\right]\Psi(z) = m^2 \Psi(z)
\ee
with a `volcano  potential' 
\be
V(z) = \frac{15\mu^2}{8(\mu |z| +1)^2} - \frac{3\mu}{2} \delta(z) ~,
\ee
which peaks as $|z|\rightarrow 0$ but has a negative 
singularity right at the origin.
It is well known from the quantum mechanics analog that such  
delta potential has a bound state, whose wave function is peaked at $z=0$, which
also means at $y=0$. In other words, there is a mode that 
appears as to be localized at the brane. 
Such a state is identified as our four dimensional graviton. Its localization is
the physical reason why gravity still behaves as four dimensional at the
brane.

Indeed, the wave function for the localized state goes as
 \be 
  \Psi_o(z) =\frac{1}{\mu (|z| +1/\mu)^{3/2}}
  \ee
whereas  KK mode wave functions in the continuum 
are written in terms of Bessel functions, in
close analogy to Eq.~(\ref{kkchi}), as
 \[ 
\Psi_m\sim s(z)
 \left[ Y_2\left( m|z| + \frac{1}{\mu}\right)
 + \frac{4\mu^2}{\pi m^2} J_2\left( m|z| + \frac{1}{\mu}\right) \right]
 \]
where $s(z) = (|z| +1/\mu)^{1/2}$. By properly normalizing these wave functions
using the asymptotic for of  Bessel functions, 
it is possible to show that 
for $m<\mu$ the wave function at brane has the value
 \be 
 h_m(0) \approx \sqrt{\frac{m}{\mu}}~.
 \ee
The coupling of gravitons to the brane is therefore weak for the lightest
KK graviton states. The volcano potential acts as a barrier for those modes. 
The production of gravitons at low energies would then be negligible.

The immediate application of our last calculations 
is on the estimation of the effective gravitational 
interaction law at the brane. The reader should remember that the 
effective interaction of brane matter to gravitons goes as
$h_{\mu\nu}(0)T^{\mu\nu}$. So, it does involve 
the evaluation of the graviton wave
function at the brane position. 
Graviton exchange  between  
two test particles on the brane separated by a distance $r$ 
then gives the effective potential~\cite{rs2grav} 
 \be
 U_{RSII}(r)\approx U_N(r)\left[1 + 
 \int_0^\infty\!\frac{dm}{\mu}\, \frac{m}{\mu} e^{-mr}\right] = 
 U_N(r)\left[1 +\frac{1}{\mu^2r^2}\right]~.
 \ee
Notice that the correction looks exactly as in the two extra dimensional ADD
case, with  $1/\mu$ as the effective size of the extra
dimensions. Thus, to the brane  the bulk should appear as compact,
at least from the gravitational point of view. The conclusion is striking.
There could be non compact extra dimensions and yet scape to our observations!.

\subsection{Beyond RSII: More Infinite Extra Dimensions }

The RSII model, which  provides a serious alternative to compactification, 
can immediately be extended to have a larger number of dimensions. 
First, notice that the metric (\ref{rsmetric})  came out of the peculiar
properties of co-dimension one objects in gravity. Thus, it is obvious that the
straightforward generalization should also contain some co-dimension one branes
in the configuration. Our brane, however should have a larger co-dimension.
Lets consider a system of $n$ mutually intersecting 
$(2+n)$ branes in a $(4+n)$
dimensional AdS space, of cosmological constant $-\Lambda$. 
All branes should have a positive tension $\tau$.
Clearly, the branes intersection is a 4 dimensional 
brane, where we assume our Universe lives. 
Intuitively, each of the $(2+n)$
branes would try to localize the graviton to itself, just as the RSII brane
does. As a consequence, the zero mode graviton would be localized at the
intersection of all branes. This naive observation can indeed be confirmed by
solving the Einstein equations for the action~\cite{nimars}
 \be
 S = \int\!d^4x\, d^n y\, \sqrt{|g_{(4+n)}|} 
 \left( \frac{1}{2k_\ast} R_{(4+n)}
 +\Lambda\right) 
-\sum_{\mbox{all branes}} \tau\,\int\!d^4x \, d^{n-1}y\,
 \sqrt{|g_{(3+n)}|} 
 \ee
If the branes are all orthogonal to each other, 
it is straightforward to see that  the space consist of $2^n$ equivalent 
slices of AdS space, glued together along the flat branes. 
The metric, therefore,  would be conformally flat. Thus, one can write it down
using appropriate bulk coordinates as
 \be   
 ds_{(4+n)}^2 = \Omega(z) \left(\eta_{\mu\nu} dx^\mu dx^\nu - 
 \delta_{kl}\, dz^k dz^l\right)
 \label{nrsmetric}
 \ee
with the warp factor $\Omega(z) = (\mu \sum_j |z^i| +1)^{-1}$;
where the $\mu$ curvature parameter is now
 \be 
 \mu^2 = \frac{2 k_\ast^2 \Lambda}{n(n+2)(n+3)};
 \label{nrsmu}
 \ee
which is a generalization of the relation given in Eq.~(\ref{rsmu}).
Similarly, the fine tuning condition (\ref{rsfine}), now  looks as
 \be 
 \Lambda = \frac{n(n+3)}{8(n+2)}\tau^2k_\ast^2~.
 \label{nrsfine}
 \ee
Effective Planck scale is now calculated to be 
 \be 
 M_P^2 = M_\ast^{(n+2)}\int\!d^n z\Omega^{(2+n)}= 
 \frac{2^nn^{n/2}}{(n+1)!} M_\ast^{(n+2)} L^n~,
 \ee
for $L= 1/\sqrt{n}\mu$. Notice this expression resembles the ADD relationship
given in Eq.~(\ref{mp}), with the effective size of the extra dimensions
proportional to $L$.

Graviton localization can be now seen by perturbing the metric with 
$\eta_{\mu\nu}\rightarrow \eta_{\mu\nu} +h_{\mu\nu}$ in Eq.~(\ref{nrsmetric}),
and writing down the equation of motion for $h_{\mu\nu}$ in the gauge
${h^\mu}_\mu=0=\partial_\mu h^{\mu\nu}$, and in conformal coordinates, to get
for $\Psi= \Omega^{(n+2)/2} h$ the linearized equation
 \be 
 \left[ -\frac{1}{2} m^2 +\left( -\frac{1}{2} \nabla_{z}^2 +
 V(z)\right)\right] \hat \Psi =0~,
 \ee
which is again nothing but a Schr\"odinger equation 
with  the effective potential 
 \be 
  V(z) = \frac{n(n+2)(n+4) \mu^2}{8}\Omega -
  \frac{(n+2)\mu}{2}\Omega\sum_j \delta(z^j)
  \ee
Indeed, the spectrum has a massless  bound state localized around  the
intersection of all delta function potentials ($z=0$), which goes as
$\Psi_{bound}\sim \Omega^{(n+2)/2}(z)$. Since the potential falls off to zero
at large $z$, there would also be a continuum of modes. Since the height of the
potential near the origin goes as $\mu^2$, all modes with small  masses,
$m<\mu$ will have suppressed wave functions, whereas those with large masses will
be un-suppressed at the origin. Therefore, the contribution of the lightest
modes to gravitational potential for two test particles at the brane  would
again come suppressed as in the RSII case. The correction to Newton's law  goes
as~\cite{nimars}
 \be
 \Delta U(r)\sim U_N(r)\left( \frac{L}{r}\right)^n ~,
 \ee
which again  behaves as in the ADD case, mimicking the case of compact
dimensions, thought there are not so.

\section{Concluding remarks}

Along the present notes we  have introduced the reader to some 
aspects of models with extra dimensions where our Universe is constrained 
to live on a four dimensional hypersurface. The study of models with extra
dimensions
has become a fruitful industry that has involved several areas of 
theoretical physics in matter of  few years. 
It is fair to say, however, that many of the current leading directions 
of research  obey more to speculative ideas that to well established facts. 
Nevertheless, the studies on the brane world are
guided by the principle of physical and mathematical consistency, and inspired
on the  possibility of  connecting, at some point, 
the models with a more fundamental theory, perhaps String
theory from where the idea of extra dimensions and branes had been borrowed.
Further motivation also comes from the possibility of experimentally 
testing these ideas  within the near future, 
something that was just unthinkable in many old models where
the fundamental gravity scale was the Planck scale. 

It is hard to address the too many interesting  
topics of the area in detail, as we would have liked, 
without facing  trouble with the limiting space of this  short notes. 
In exchange, we have concentrated the discussion to the construction of  
the main  frameworks (ADD, and RS models), paying special attention to 
dimensional reduction  and some of the interesting phenomenology of 
quantum gravity  in colliders as well as in BBN and astrophysics. 
We have also discussed at some extend the calculation of the effective 
gravity interactions on the brane. Some  applications and uses in model building
for the extra dimensions have also been addressed, including the promotion of SM
fields to the bulk and the consequent  power law
running of couplings as well as some ideas on symmetry breaking
with extra dimensions. 
We have also commented some ideas 
to get a small neutrino mass and a very stable proton despite the possibility of
having a low fundamental scale. The basis of RS models have also been 
discussed in some detail.

I hope these notes could serve the propose of been a brief  introduction
to this area of research.
The list of topics we have been unable to cover is extensive, 
it includes many issues  on cosmology of models with flat and warped extra 
dimensions~\cite{cosmo1,cosmo2};
the  discussions on the cosmological constant problem~\cite{cc},
higher dimensional warped spaces~\cite{hdw}; 
dark matter from KK modes~\cite{dmkk}, many ideas on 
Black Holes in both ADD and RS models~\cite{bh1,rsbh}, 
deconstruction of extra dimensions~\cite{deconstruction}, and the list go on.
I hope the  interested readers could consult other 
reviews~\cite{reviews} and  search  for further references.

\acknowledgments
The author's work is supported in part by CONACyT Mexico, 
under grant J44596-F.


\end{document}